\documentclass[aps,prd,reprint,superscriptaddress,twocolumn,showpacs,floatfix,10pt]{revtex4-1}
\usepackage{color}
\usepackage{hyperref}
\usepackage{enumerate}
\usepackage{amsmath}
\usepackage{graphicx}
\usepackage{subfigure}

\def \be {\begin{equation}}
\def \ee {\end{equation}}
\def \ba {\begin{array}}
\def \ea {\end{array}}
\def \bea {\begin{eqnarray}}
\def \eea {\end{eqnarray}}

\def \ble {\begin{widetext}\begin{equation}}
\def \ele {\end{equation}\end{widetext}}
\def \blea {\begin{widetext}\begin{eqnarray}}
\def \elea {\end{eqnarray}\end{widetext}}

\def \nn {\nonumber}

\def \blea {\begin{widetext}\begin{eqnarray}}
\def \elea {\end{eqnarray}\end{widetext}}
\def \mO {\mathcal{O}}

\def \mA {\mathcal{A}}

\def \be {\begin{equation}}
\def \ee {\end{equation}}
\def \ba {\begin{array}}
\def \ea {\end{array}}
\def \bea {\begin{eqnarray}}
\def \eea {\end{eqnarray}}
\def \nn {\nonumber}

\def \p {\partial}

\def \and {{~\textrm{and}~}}

\begin{document}

\title{Relation between time- and spacelike entanglement entropy}



\author{Wu-zhong Guo}
\email{Corresponding author:wuzhong@hust.edu.cn}
\affiliation{School of Physics, Huazhong University of Science and Technology,
Luoyu Road 1037, Wuhan, Hubei 430074, China}

\author{Song He}
\email{Corresponding author:hesong@jlu.edu.cn}
\affiliation{Institute of Fundamental Physics and Quantum Technology, Ningbo University,\\  Ningbo, 315211, China}
\affiliation{School of Physical Science and Technology, Ningbo University,\\
Ningbo, 315211,  China}
\affiliation{Center for Theoretical Physics and College of Physics, Jilin University,\\
Changchun 130012, China}

\author{Yu-Xuan Zhang}
\email{Corresponding author:yuxuanz20@mails.jlu.edu.cn}
\affiliation{Center for Theoretical Physics and College of Physics, Jilin University,\\
Changchun 130012, China}

\date{}



\begin{abstract}
In this study, we establish a connection between timelike and spacelike entanglement entropy. We show that timelike entanglement entropy is closely related to spacelike entanglement entropy and its temporal derivative. For a broad class of states, it can be uniquely determined by a linear combination of spacelike entanglement entropy and its first-order temporal derivative. This relation holds, for instance, in states conformally equivalent to the vacuum in two-dimensional conformal field theories. For more general states, we demonstrate that the relation can be constructed perturbatively. 
Our results suggest that timelike entanglement entropy is constrained by causality. Moreover, this relation provides a unified framework for timelike and spacelike entanglement entropy, within which the imaginary component of timelike entanglement entropy can be understood as arising from the non-commutativity between the twist operator and its first-order temporal derivative.
\end{abstract}

\date{}
\maketitle

\section{Introduction}

%



In the realm of quantum field theory (QFT) and quantum many-body systems, entanglement entropy (EE) has become a pivotal tool for probing quantum correlations. Traditionally, much focus has been on spatial separations, with extensive research significantly advancing our understanding of EE \cite{Amico:2007ag, Calabrese:2009qy, Rangamani:2016dms}. In theories with holographic duals, the connection between EE and minimal surfaces within the gravitational context has been notably established by the Ryu-Takayanagi (RT) formula \cite{Ryu:2006bv} and its extension, the Hubeny-Rangamani-Takayanagi (HRT) formula \cite{Hubeny:2007xt}, proving powerful in probing quantum black holes and quantum gravity \cite{VanRaamsdonk:2010pw, Maldacena:2013xja, Almheiri:2014lwa, Penington:2019npb, Almheiri:2019psf}.



The concept of EE is typically defined for a subsystem that constitutes a spacelike region. Recently, it has been extended to timelike regions \cite{Doi:2022iyj} in QFTs, referred to as timelike entanglement entropy, which is generally complex-valued. A proper definition of timelike EE involves the analytical continuation of the result of EE to timelike case. It appears that this new quantity diverges from previous attempts to explore the entanglement linked with time from other aspects\cite{Leggett:1985zz,Fitzsimons:2013gga,Olson:2010jy,Giudice:2021smd,Liu:2022ugc}. It is proposed in \cite{Doi:2022iyj} that timelike EE should be interpreted as the so-called pseudoentropy \cite{Nakata:2020luh}, also discussed in \cite{Murciano:2021dga}. Pseudoentropy is defined by replacing the reduced density matrix $\rho_A$ with a non-Hermitian transition matrix, leading to a generally complex-valued entropy. For further advancements in this area, refer to \cite{Mollabashi:2020yie}-\cite{Gong:2025pnu}.

In \cite{Doi:2022iyj}, the timelike entanglement entropy (EE) was defined for a time interval in $1+1$D QFTs by exchanging the roles of time and space. The result in two-dimensional CFTs agrees with the holographic expectation \cite{Doi:2023zaf}. However, this approach is difficult to generalize to higher-dimensional QFTs, since the numbers of spatial and temporal coordinates are not the same. More recently, \cite{Milekhin:2025ycm} proposed a clearer definition of entanglement in time by generalizing the density matrix to causally connected subsystems. In QFTs, such a generalized density matrix can be prepared using the Schwinger–Keldysh path integral, allowing one to compute the timelike EE via the replica method. See also \cite{Gong:2025pnu} for further discussion of entanglement measures in QFTs. Therefore, the concept of timelike EE can be well defined through the generalized density matrix. In two-dimensional CFTs, the evaluation of timelike EE reduces to correlation functions of twist operators in Lorentzian signature, reproducing the results obtained in \cite{Doi:2023zaf} by analytic continuation.

In QFTs, the most fundamental quantities are the correlation functions of local operators. The spacelike correlation functions reflect quantum correlations or entanglement between two causally disconnected regions. We also consider correlation functions for operators with timelike separations in QFTs, which are generally related to the causality and dynamics of the underlying theories. Additionally, one can characterize the entanglement between spacelike regions using various entanglement measures, with EE being the most useful, as mentioned earlier.

In principle, the evaluation of EE can be translated into computing the correlation functions of local operators in Euclidean QFTs. There exists a standard method to obtain Lorentzian correlation functions through the analytical continuation of their Euclidean counterparts. Therefore, from this perspective, it is natural to define the timelike entanglement in QFTs as the result of analytical continuation of Euclidean correlators. In the following sections, we will adopt this approach to define the timelike EE as the analytical continuation of Euclidean correlators.

Despite notable progress in this field, there remains a significant gap in our comprehensive understanding of the physical significance of timelike EE. A question arises regarding its interpretation as a concept of entropy, especially given its general complex-valued nature. A crucial objective is to understand the origin of the imaginary component of timelike EE. While it is possible to establish timelike EE through the analytical continuation of spacelike EE, the extent of the intrinsic relationship between these two aspects of entanglement entropy remains unclear.
This paper aims to construct and explore the intricate relationship between these two facets of EE. We will demonstrate that timelike and spacelike EE can be approached within a unified framework, and that timelike EE can be linked to spacelike EE and its time derivative on the Cauchy surface.



The paper is organized as follows: In Section \ref{Analytical_continuation_section}, we discuss how to obtain the timelike EE via the analytical continuation of Euclidean correlators. In Section.~ \ref{TEE_SEE_relation_section}, we present the relationship between timelike and spacelike EE for the vacuum state (\ref{linearcombination}), including a proof leveraging the similarity between R\'enyi entropy and correlators of a free scalar field. In Section.~\ref{Sec_General_section}, we demonstrate that this relation holds for thermal states and states dual to pure AdS$_3$. We also analyze the imaginary part of the timelike EE and discuss the significance of the relation, which aids in understanding the transition matrix associated with the timelike EE. In Section.~ \ref{General_section}, we further explore the most general states using the operator product expansion (OPE) of twist operators.  Finally, Section \ref{discuss_section} concludes the paper with a discussion of our findings.

\section{Timelike entanglement entropy by analytical continuation}\label{Analytical_continuation_section}

The system is characterized by the density matrix $\rho$, defining the reduced density matrix as $\rho_A:=\text{tr}_{\bar A}\rho$. EE is given by the von Neumann entropy $S(\rho_A)=-\text{tr}(\rho_A\log\rho_A)$. To evaluate it, we introduce the Rényi entropy (RE) $S_n(\rho_A)=\frac{\log\text{tr}(\rho_A^n)}{1-n}$. In QFTs, we use the replica method to compute RE, leading to the twist operator formalism \cite{Calabrese:2004eu}. More precisely, preparing the state $\rho=|\psi\rangle\langle \psi|$ by Euclidean path integral, the $n$-th Rényi entropy can be computed by a path integral for $n$-copied systems glued together along the subsystem $A$, denoted as $\Sigma_n$. Let us focus on 2-dimensional conformal field theories (CFTs). If $A$ is an interval, $tr(\rho_A)^n$ can be expressed as a correlator involving twist operators for the $n$-copied CFT$_n$, that is
\bea\label{twist}
tr(\rho_A)^n=\langle \Psi|\sigma_n(\tau,x)\tilde{\sigma}_n(\tau',x')|\Psi\rangle,
\eea
where $(\tau,x)$ and $(\tau',x')$ are coordinates of the endpoints of $A$,  $|\Psi\rangle:=|\psi\rangle_1\otimes ...|\psi\rangle_i \otimes ... |\psi\rangle_n$, the subscripts $i$ label the $i$-th copy. In 2-dimensional CFTs, the twist operators can be taken as local primary operators with the conformal dimension $h_n=\bar h_n=\frac{c}{24}(n-\frac{1}{n})$.

The twist operator formalism of EE provides a natural framework for extending the concept of entanglement into the timelike region. Spacelike entanglement captures the quantum correlations between different spacelike regions. While the correlators of local operators are the most fundamental quantity used to characterize these correlations. In QFTs, it is essential to consider the timelike correlators of two local operators, as these are intrinsically linked to causality. In two-dimensional QFTs, EE can actually be evaluated using the correlators of twist operators. From this perspective, we argue that it is appropriate to define timelike EE through the use of twist operators. In the replica approach to calculating EE, one typically works within Euclidean QFTs. Furthermore, there exists a well-established method in QFTs to obtain Lorentzian correlators through analytic continuation of their Euclidean counterparts.

We would like to use the vacuum state as an example to demonstrate how to obtain the timelike entanglement entropy through analytical continuation. Consider the vacuum state $|\psi\rangle=|0\rangle$, where
\begin{equation}
    \begin{aligned}
        &\text{tr}(\rho_{0,A})^n:=\langle \sigma_n(\tau,x)\tilde{\sigma}_n(\tau',x')\rangle \\
        &\phantom{\text{tr}(\rho_{0,A})^n:}=(w_1-w_2)^{-2h_n}(\bar w_1-\bar w_2)^{-2\bar h_n},
    \end{aligned}
\end{equation}
with $w_i=x_i+i\tau_i$ ($i=1,2$).

We are interested in an interval between $(t,x)$ and $(t',x')$ in Minkowski spacetime with these two points being timelike. The timelike R\'enyi entropy can be obtained by Wick rotation $\tau\to it+\epsilon$. More generally, for any interval between $(t,x)$ and $(t',x')$, we have
\begin{equation}\label{Snvacuum}
    \begin{aligned}
        &S_n(t,x;t',x'):= \frac{\log \text{tr}(\rho_{0,A})^n|_{\tau\to it+\epsilon,\ \tau'\to it'+\epsilon'} }{1-n} \\
        &=\frac{2h_n}{n-1}\log \left[ \frac{\Delta s^2+2i(\epsilon-\epsilon')(t-t')}{\delta^2} \right],
    \end{aligned}
\end{equation}
where $\Delta s^2=-(t-t')^2+(x-x')^2$ and $\epsilon>\epsilon'$, $\delta$ is the UV cut-off.

For spacelike separation $\Delta s^2>0$, $S_n(t,x;t',x')$ is the spacelike R\'enyi entropy, satisfying the symmetry $S_n(t,x;t',x')=S_n(t',x';t,x)$. For timelike separation with $t>t'$, where $\Delta s^2<0$, we have
\begin{equation}\label{timelike_Renyi_vacuum}
    S_n(t,x;t',x')=\frac{c}{12}\left(1+\frac{1}{n}\right)\log [-\frac{\Delta s^2}{\delta^2}]+\frac{i\pi c}{12}\left(1+\frac{1}{n}\right).
\end{equation}
In this case, $S_n(t',x';t,x)=S_n(t,x;t',x')^*$, and the imaginary part is given by
\begin{equation}
    \begin{aligned}
        \text{Im}[S_n(t,x;t',x')] &= \frac{1}{2}\left[S_n(t,x;t',x')-S_n(t',x';t,x)\right] \\
        &=\frac{i\pi c}{12}\left(1+\frac{1}{n}\right).
    \end{aligned}
\end{equation}
Taking the limit $n\to 1$ yields the timelike EE. This gives the correct timelike EE as one done in \cite{Doi:2022iyj}, where the authors define the timelike EE by exchanging the role of spatial and temporal coordinate. 

\section{Timelike and spacelike EE relation in vacuum state}\label{TEE_SEE_relation_section}

From the perspective of correlators, it is natural to establish a connection between timelike correlators and spacelike ones. For a theory with Hamiltonian $H$, the dynamics of a local operator $O(t,\vec{x})$ is given by $O(t,\vec{x}):=e^{iH t}O(0,\vec{x})e^{-iHt}$, which can be expressed formally as 
\bea\label{evolution}
O(t,\vec{x})=\sum_m \frac{(it)^m}{m!}[H,...,[H,O(0,\vec{x})]],
\eea
which means the operator $O(t,\vec{x})$ can be expressed as linear combinations of operators $[H,...,[H,O(0,\vec{x})]]$ on the Cauchy surface at $t=0$. Similarly, for another operator $O'(t',\vec{x}')$, a similar decomposition can be performed. Consequently, the correlator $\langle O(t,\vec{x})O'(t',\vec{x}')\rangle$ can be formally represented as linear combinations of spacelike correlators. However, in general, the expression becomes too complex to be practically useful.

Timelike EE can be more properly understood in 2-dimenional CFTs. In this context, the evaluation of EE can be translated to the correlation functions of the local twist operator $\sigma_n$ in the $n$-copied CFT$_n$ theory\cite{Calabrese:2004eu}. One can also define its time evolution by the Hamiltonian $H^{(n)}$ of CFT$_n$ theory. The timelike EE can be computed through the analytical continuation from Euclidean to Lorentzian correlation function of twist operators.
By utilizing (\ref{evolution}) for the twist operator, one anticipates that the timelike EE can be linked to correlators on the Cauchy surface at $t=0$. However, our findings extend beyond this expectation. We will demonstrate in various cases that the timelike EE is equivalent to a linear combination of \textit{solely} spacelike EE and its first-order time derivative.

\begin{figure}
\centering
\subfigure[]{\includegraphics[scale=0.4]{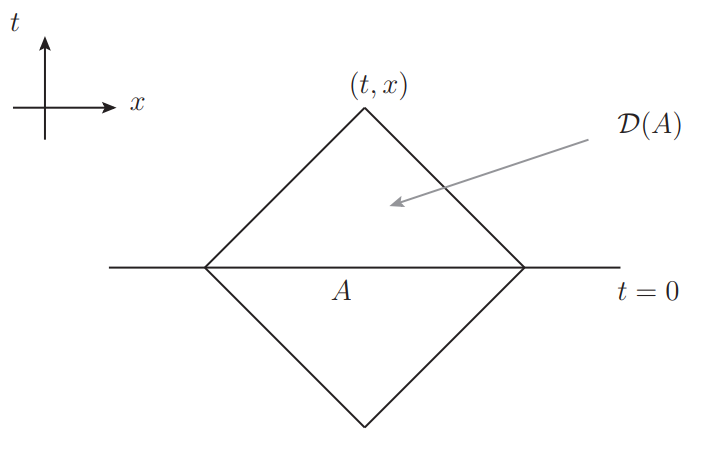}}
\subfigure[]{\includegraphics[scale=0.4]{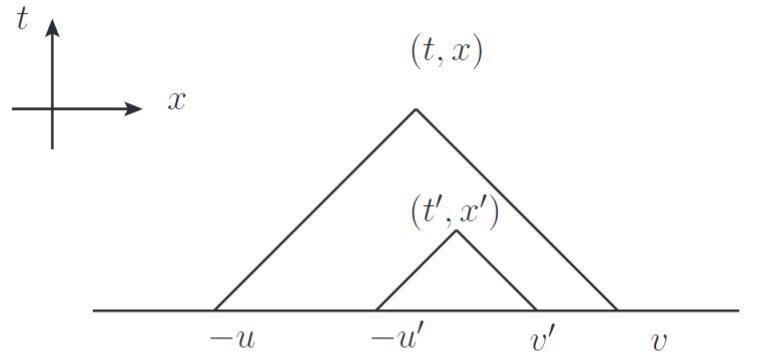}}
\caption{(a) An subregion $A$ on the Cauchy surface $t=0$ and its causal domain of dependence $\mathcal{D}(A)$. (b) In a typical scenario where $(t,x)$ and $(t',x')$ are timelike, their past light cones intersect with four points at $t=0$ with $-u<-u'<v'<v$. }
\label{CausalDomain}
\end{figure}

  For an open subregion $A$ on the Cauchy surface $t=0$ in a $d$-dimensional spacetime, the local algebra $\mathcal{R}(A)$ is constructed using smeared local operators in $A$. The domain of dependence of $A$, denoted by $\mathcal{D}(A)$, is the set of spacetime points $(t,\vec{x})$ causally influenced by or influencing points in $A$ (refer to Fig.\ref{CausalDomain}). The dynamical time evolution of the theory determines operators in $\mathcal{D}(A)$ by those in the region $A$. Thus, one expects the algebra $\mathcal{R}[\mathcal{D}(A)]$ associated with $\mathcal{D}(A)$ to be equal to the algebra $\mathcal{R}(A)$. However, the local operator $\mathcal{O}(t,\vec{x})$ in $\mathcal{R}(\mathcal{D}_A)$ generally has a complex relation with those $\{ \mathcal{O}_i(0,\vec{y})\}$ in $\mathcal{R}(A)$, expressed as
\bea\label{formrelation}
\mO(t,\vec{x})=\sum_i\int_{A}d^{d-1}\vec{y}f_i(t,\vec{x};0,\vec{y}) \mO_i(0,\vec{y}),
\eea 
where $f_i(t,\vec{x};0,\vec{y})$ are functions depending on the coordinates. For a general theory, Eq. (\ref{formrelation}) is seen as a form relation. 

\subsection{Derivation of the relation}

In this section we will focus on vacuum state and derive a relation between timelike and spacelike EE. The relation can be expressed as 
\begin{widetext}
\bea\label{linearcombination}
&&S(t,x;t',x')=\frac{1}{4}\Big(S(0,-u;0,-u')+S(0,-u;0,v')+S(0,v;0,-u')+S(0,v;0,v') \Big)\nn \\
&&\phantom{S(t,x;t',x')}+\frac{1}{4}\int_{-u'}^{v'} d\bar x'\partial_{t'}S(0,-u;0,\bar x')+\frac{1}{4}\int_{-u'}^{v'} d\bar x'\partial_{t'}S(0,v;0,\bar x')\rangle\nn \\
&&\phantom{S(t,x;t',x')}+\frac{1}{4}\int_{-u}^{v} d\bar x \partial_{t}S(0,\bar x;0,-u')+\frac{1}{4}\int_{-u}^{v} d\bar x \partial_t S(0,\bar x;0,v')\nn \\
&&\phantom{S(t,x;t',x')}+\frac{1}{4}\int_{-u}^{v} d\bar x \int_{-u'}^{v'} d\bar x' \partial_t\partial_{t'}S(0,\bar x;0,\bar x'),
\eea
\end{widetext}
where we define $u=t-x$, $v=t+x$ and $u'=t'-x'$, $v'=t'+x'$.
The time derivative term should be interpreted as obtaining the derivative first and then taking the limit as  $t\to 0$. For example, $\partial_{t'}S(0,-u;0,\bar x')$ should be understood as $\partial_{t'}S(0,-u;t',\bar x')|_{t'\to 0}$. Actually for the vacuum state the above equation is also correct for the RE with replacing $S(t,x;t',x')$ by $S_n(t,x;t',x')$. This is the main result of our paper. 

To prove the formula in Eq. (\ref{linearcombination}), we will utilize the similarity between the Rényi entropy in the vacuum state and the correlators of $\phi$ in a 2-dimensional massless free scalar theory. The rough process is as follows. Given the action of massless free scalar theory $$S=\frac{1}{8\pi \kappa}\int dt dx [(\p_t\phi)^2-(\p_x\phi)^2].$$ 
The operator $\phi(t,x)$ can be expressed as linear combinations of the operators $\phi(0,x)$ and $\pi(0,x):=\frac{1}{4\pi \kappa}\dot{\phi}$ located in region $A$ . Thus the timelike correlator $\langle  \phi(t,x) \phi(t',x')\rangle$ can be written as combinations of  spacelike correlators $\langle \phi(0,x)\phi(0,x')\rangle $, $\langle \phi(0,x)\pi(0,x')\rangle$ and $\langle \pi(0,x)\pi(0,x')\rangle$. A noteworthy observation is that the correlator $\langle \phi(t,x) \phi(t',x')\rangle$ coincides with the R\'enyi entropy in the vacuum (\ref{Snvacuum}) when replacing $\kappa$ with $-\frac{2h_n}{n-1}$.  Hence, we can derive an intriguing relation that links timelike EE with spacelike EE and their first-order time derivative with the help of the correlators in massless free scalar theory. 

\subsection{Proof of the relation for vacuum state}
In Appendix.\ref{app_A}, we review some details of the correlators in massless free theory for the field  $\phi$. For the operator $\phi(t,x)$, it can be expressed in terms of operators situated on the Cauchy surface at $t=0$. In other words, we could derive the relationship given in Eq. (\ref{formrelation}) for $\phi(t,x)$. For $\phi(t,x)$ we have the relation
\bea\label{phirelation}
\phi(t,x)=\int_{A} d\bar x f_\phi(t,x;0,\bar x)\phi(0,\bar x)+\int_{A} d\bar x f_\pi(t,x;0,\bar x)\pi(0,\bar x),\nn
\eea
where $\pi(t,x):=\frac{1}{4\pi \kappa}\dot{\phi}$ is the canonical momentum operator, and the functions $f_{\phi(\pi)}(t,\vec{x};0,\vec{\bar x})$ are given by
\bea\label{ff_main}
&&f_{\phi}(t,x;t',x')
=\frac{1}{2} \left(\delta(u-u')+\delta(v-v') \right),\nn 
\\
&&f_{\pi}(t,x;t',x')
=\pi\kappa\left(H(u-u')-H(v'-v)\right),\label{fphifpi}
\eea
where $u:=t-x$, $v=t+x$ and $u'=t'-x'$, $v'=t'+x'$, $H(x)$ is Heaviside function. Refer to Appendix.\ref{app_A} for the derivation of the above formula. The formula can also be expressed in a more explicit form as follows
\bea\label{operatorsumrule_phi}
\phi(t,x)=\frac{1}{2}\phi(0,-u)+\frac{1}{2}\phi(0,v)+2\pi \kappa \int_{-u}^{v}d\bar x \pi(0,\bar x).\nn \\
~
\eea

Thus the timelike correlator $\langle \phi(t,x)\phi(t',x')\rangle$ can be expanded as linear combinations of the spacelike correlators, that is
\begin{widetext}
\bea
&&\langle \phi(t,x)\phi(t',x')\rangle=\int_{\Delta}d\bar x d\bar x' f_\phi(t,x;0,\bar x)f_\phi(t',x';0,\bar x')\langle \phi(0,\bar x)\phi(0,\bar x')\rangle\nn \\
&&\phantom{phi(t,x)\phi(t',x')\rangle=}+\int_{\Delta}d\bar x d\bar x' f_\phi(t,x;0,\bar x)f_\pi(t',x';0,\bar x')\langle \phi(0,\bar x)\pi(0,\bar x')\rangle\nn \\
&&\phantom{phi(t,x)\phi(t',x')\rangle=}+\int_{\Delta}d\bar x d\bar x' f_\pi(t,x;0,\bar x)f_\phi(t',x';0,\bar x')\langle \pi(0,\bar x)\phi(0,\bar x')\rangle\nn \\
&&\phantom{phi(t,x)\phi(t',x')\rangle=}+\int_{\Delta}d\bar x d\bar x' f_\pi(t,x;0,\bar x)f_\pi(t',x';0,\bar x')\langle \pi(0,\bar x)\pi(0,\bar x')\rangle.\nn
\eea

Using the result (\ref{phirelation}), we have 
\bea\label{timecorrelator_sum}
&&\langle \phi(t,x)\phi(t',x')\rangle\nn \\
&&=\frac{1}{4}\left(\langle \phi(0,-u)\phi(0,-u')\rangle+\langle \phi(0,-u)\phi(0,v')\rangle+\langle \phi(0,v)\phi(0,-u')\rangle+\langle \phi(0,v)\phi(0,v')\rangle \right)\nn \\
&&+\pi \kappa\int_{-u'}^{v'} d\bar x'\langle \phi(0,-u)\pi(0,\bar x')\rangle+\pi \kappa\int_{-u'}^{v'} d\bar x'\langle \phi(0,v)\pi(0,\bar x')\rangle\nn \\
&&+\pi \kappa\int_{-u}^{v} d\bar x \langle \pi(0,\bar x)\phi(0,-u')\rangle+\pi \kappa\int_{-u}^{v} d\bar x \langle \pi(0,\bar x)\phi(0,v')\rangle\nn \\
&&+4(\pi \kappa)^2\int_{-u}^{v} d\bar x \int_{-u'}^{v'} d\bar x' \langle \pi(0,\bar x)\pi(0,\bar x')\rangle,
\eea
\end{widetext}
where $u=t-x$, $v=t+x$ and $u'=t'-x'$, $v'=t'+x'$.

Eq.(\ref{timecorrelator_sum}) shows the timelike correlator $\langle \phi(t,x)\phi(t',x')$ can be expressed as linear combinations of spacelike correlators.  We can verify the equation (\ref{timecorrelator_sum}) directly. Each term on the right-hand side of (\ref{timecorrelator_sum}) can be evaluated as follows.
Using (\ref{phiphicorrelator}) we have
\bea
&&\frac{1}{4}\big(\langle \phi(0,-u)\phi(0,-u')\rangle+\langle \phi(0,-u)\phi(0,v')\rangle\nn \\
&&+\langle \phi(0,v)\phi(0,-u')\rangle+\langle \phi(0,v)\phi(0,v')\rangle big)\nn \\
&&=-\frac{\kappa}{4}\big(\log(u-u')^2+\log(u-v')^2\nn \\
&&+\log(v+u')^2\nn +\log(v-v')^2 \big).
\eea
Using (\ref{correlatorphipi_canonical}) we have
\bea
&&\pi \kappa\int_{-u'}^{v'} d\bar x'\langle \phi(0,-u)\pi(0,\bar x')\rangle\nn \\
&&+\pi \kappa\int_{-u'}^{v'} d\bar x'\langle \phi(0,v)\pi(0,\bar x')\rangle\rangle\nn \\
&&=\frac{\kappa}{4}\int_{-u'}^{v'}d\bar x' \left(\frac{1}{u+\bar x'-i\epsilon}+\frac{1}{-u-\bar x'-i \epsilon} \right)\nn \\
&&+\frac{\kappa}{4}\int_{-u'}^{v'}d\bar x' \left(\frac{1}{-v+\bar x'-i\epsilon}+\frac{1}{v-\bar x'-i \epsilon} \right)\nn \\
&&=0,
\eea
where we use the fact $-u<-u'<v'<v$. Similarly, we obtain 
\bea
&&\pi \kappa\int_{-u}^{v} d\bar x \langle \pi(0,\bar x)\phi(0,-u')\rangle\nn \\
&&+\pi \kappa\int_{-u}^{v} d\bar x \langle \pi(0,\bar x)\phi(0,v')\rangle\nn \\
&&= \frac{\kappa}{4}\int_{-u}^{v} d\bar x \left(\frac{1}{-\bar x-u'-i\epsilon}+\frac{1}{\bar x+u'-i\epsilon} \right)\nn \\
&&+\frac{\kappa}{4}\int_{-u}^{v} d\bar x \left(\frac{1}{-\bar x+v'-i\epsilon}+\frac{1}{\bar x-v'-i\epsilon} \right)\nn \\
&&=i\pi\kappa,
\eea

where we use the Sokhotski–Plemelj theorem
\bea
\frac{1}{x-i\epsilon}=i \pi\delta(x)+\mathcal{P}\left( \frac{1}{x}\right),
\eea
where $\mathcal{P}$  denotes the Cauchy principal value. The final term is
\begin{widetext}
\bea\label{ptptS}
&&4(\pi \kappa)^2\int_{-u}^{v} d\bar x \int_{-u'}^{v'} d\bar x' \langle \pi(0,\bar x)\pi(0,\bar x')\rangle\nn \\
&&=-\frac{\kappa}{4}\int_{-u}^vd\bar x \int_{-u'}^{v'}d\bar x' \left[\frac{1}{(\bar x-\bar x'+i \epsilon)^2}+\frac{1}{(\bar x-\bar x' -i\epsilon)} \right]\nn \\
&&=-\frac{\kappa}{4}\int_{-u}^vd\bar x \int_{-u'}^{v'}d\bar x' \p_{\bar x'}\left(\frac{1}{\bar x-\bar x'+i\epsilon}+\frac{1}{\bar x-\bar x'-i\epsilon} \right)\nn \\
&&=-\frac{\kappa}{4}\int_{-u}^vd\bar x \left(\frac{1}{\bar x-v'+i\epsilon}-\frac{1}{\bar x+u'+i\epsilon}+\frac{1}{\bar x-v'-i\epsilon}-\frac{1}{\bar x+u'-i\epsilon} \right)\nn \\
&&=\frac{\kappa}{2}\log \frac{u+v'}{v-v'}+\frac{\kappa}{2}\log \frac{u'+v}{u-u'},
\eea
\end{widetext}
where in the last step we use the Sokhotski–Plemelj theorem again.
Summing over all the terms we find the result is
\bea\label{timelike_correlator_phi}
-\kappa\log \left[(u-u')(v-v')\right]+i\pi \kappa,
\eea
which is consistent with (\ref{phiphicorrelator}).

Comparing Eq.(\ref{timelike_Renyi_vacuum}) and Eq.(\ref{timelike_correlator_phi}), it is interesting to note that $S_n(t,x;t',x')$ in the vacuum state is same as the two point correlator $\langle  \phi(t,x) \phi(t',x')\rangle$ with $\kappa\to -\frac{2h_n}{n-1} $. One can also show that
\bea
&&\frac{1-n}{8\pi h_n}\partial_{t'}S_n(t,x;t',x')=\langle \phi(t,x)\pi(t',x')\rangle,\nn \\
&&\left(\frac{1-n}{8\pi h_n}\right)^2 \partial_t\partial_{t'}S_n(t,x;t',x')=\langle \pi(t,x)\pi(t',x')\rangle.
\eea
Using the Eq.(\ref{timecorrelator_sum}) and the above formulae, we can derive a relation between  timelike R\'enyi entropy and spacelike R\'enyi entropy,
\begin{widetext}
\bea\label{linearcombinationalt}
&&S_n(t,x;t',x')=\frac{1}{4}\Big(S_n(0,-u;0,-u')+S_n(0,-u;0,v')+S_n(0,v;0,-u')+S_n(0,v;0,v') \Big)\nn \\
&&\phantom{S_n(t,x;t',x')}+\frac{1}{4}\int_{-u'}^{v'} d\bar x'\partial_{t'}S_n(0,-u;0,\bar x')+\frac{1}{4}\int_{-u'}^{v'} d\bar x'\partial_{t'}S_n(0,v;0,\bar x')\rangle\nn \\
&&\phantom{S_n(t,x;t',x')}+\frac{1}{4}\int_{-u}^{v} d\bar x \partial_{t}S_n(0,\bar x;0,-u')+\frac{1}{4}\int_{-u}^{v} d\bar x \partial_t S_n(0,\bar x;0,v')\nn \\
&&\phantom{S_n(t,x;t',x')}+\frac{1}{4}\int_{-u}^{v} d\bar x \int_{-u'}^{v'} d\bar x' \partial_t\partial_{t'}S_n(0,\bar x;0,\bar x').
\eea
\end{widetext}
Taking the limit $n\to 1$ we obtain the relation for timelike entanglement entropy Eq.(\ref{linearcombination})

\section{Extenstion of the relation to more general states}\label{Sec_General_section}
Although we derived this formula for the vacuum state, one can show it is exactly correct for broad cases:
\begin{itemize}
\item One interval in thermal states with temperature $1/\beta$, or vacuum state on a cylinder.
\item One interval in the states that have holographic duals to  AdS$_3$ satisfying vacuum Einstein equation.
\item Multiple intervals in the states with holographic dual to the leading order of $O(G)$. The RT formula allows for the evaluation of EE to the leading order in $G$. The holographic EE for multiple intervals is obtained by summing those for individual intervals. Yet, different phases may emerge depending on the lengths of the intervals.  In the single interval case, we have established the validity of the relation (\ref{linearcombination}) for the states with holographic dual. Consequently, the timelike EE for multiple intervals also have similar relation.
\end{itemize}
We will show these examples in the following sections.\\

\subsection{ The relation  for thermal state}\label{Thermal_section}

In this section, we would like to show the relation (\ref{linearcombination}) is correct for the thermal state. For a CFT living on an infinite line with temperature $1/\beta$, the EE  is given by
\bea
&&S(t,x;t',x')=\frac{c}{6}\log\Big[\frac{\beta^2}{\pi^2 \delta^2}\sinh\left(\frac{\pi(v-v'-i\epsilon)}{\beta} \right)\nn \\
&&\phantom{S(t,x;t',x')=\frac{c}{6}\log}\times\sinh\left(\frac{\pi(u'-u+i\epsilon)}{\beta} \right) \Big],
\eea
where $\epsilon>0$. For $(t,x)$ and $(t',x')$ being timelike, we have $(v-v')(u-u')>0$, thus the timelike EE is
\bea\label{thermaltimelike}
&&S(t,x;t',x')=\frac{c}{6}\log\Big[\frac{\beta^2}{\pi^2 \delta^2}\sinh\left(\frac{\pi(v-v')}{\beta} \right)\nn \\
&&\phantom{S(t,x;t',x')=\frac{c}{6}}\times\sinh\left(\frac{\pi(u-u')}{\beta} \right) \Big]
+\frac{i\pi c}{6}.
\eea
The derivation of the relation (\ref{linearcombination}) in the vacuum state cannot be directly generalized to the thermal state.

The four terms without temporal derivative are given by
\begin{widetext}
\bea
&&\frac{1}{4}\left(S_n(0,-u;0,-u')+S_n(0,-u;0,v')+S_n(0,v;0,-u')+S_n(0,v;0,v') \right)\nn \\
&&=\frac{c}{48}\left(1+\frac{1}{n}\right)\Big(\log\left[\frac{\beta^2}{\pi^2 \delta^2}\sinh^2\left(\frac{\pi(u-u')}{\beta}\right) \right]+\log\left[\frac{\beta^2}{\pi^2 \delta^2}\sinh^2\left(\frac{\pi(u+v')}{\beta}\right) \right]\nn \\
&&\phantom{\frac{c}{48}\left(1+\frac{1}{n}\right)}+\log\left[\frac{\beta^2}{\pi^2 \delta^2}\sinh^2\left(\frac{\pi(u'+v)}{\beta}\right) \right]+\log\left[\frac{\beta^2}{\pi^2 \delta^2}\sinh^2\left(\frac{\pi(v-v')}{\beta}\right) \right]  \Big).\nn\\
~
\eea
\end{widetext}
The middle four terms will contribute the imaginary part of timelike EE, the results are given by
\begin{widetext}
\bea
\frac{1}{4}\int_{-u'}^{v'} d\bar x'\partial_{t'}S_{n}(0,-u;0,\bar x')+\frac{1}{4}\int_{-u'}^{v'} d\bar x'\partial_{t'}S_n(0,v;0,\bar x')\rangle=0.
\eea
\bea\label{thermalimaginary}
&&\frac{1}{4}\int_{-u}^{v} d\bar x \partial_{t}S_n(0,\bar x;0,-u')+\frac{1}{4}\int_{-u}^{v} d\bar x \partial_t S_n(0,\bar x;0,v')\nn \\
&&=\frac{c \pi}{48\beta}\left(1+\frac{1}{n} \right)\int_{-u}^v d\bar x \left(-\coth(\frac{\pi(\bar x+u'+i\epsilon)}{\beta}) +\coth(\frac{\pi(\bar x+u'-i\epsilon)}{\beta})\right)\nn\\
&&+\frac{c \pi}{48\beta}\left(1+\frac{1}{n} \right)\int_{-u}^v d\bar x \left(-\coth(\frac{\pi(\bar x-v'+i\epsilon)}{\beta}) +\coth(\frac{\pi(\bar x-v'-i\epsilon)}{\beta})\right)\nn \\
&&=\frac{i\pi c}{12}\left(1+\frac{1}{n} \right).
\eea
\end{widetext}
The final term is real and given by
\begin{widetext}
\bea
&&\frac{1}{4}\int_{-u}^{v} d\bar x \int_{-u'}^{v'} d\bar x' \partial_t\partial_{t'}S_n(0,\bar x;0,\bar x')\nn \\
&&=\frac{c \pi^2}{48\beta^2}\left(1+\frac{1}{n} \right)\int_{-u}^v d\bar x \int_{-u'}^{v'}d\bar x' \left[\frac{1}{\sinh^2 \left(\frac{\pi(\bar x-\bar x'-i\epsilon)}{\beta}\right)}+\frac{1}{\sinh^2 \left(\frac{\pi(\bar x-\bar x'+i\epsilon)}{\beta}\right)}\right]\nn \\
&&=\frac{c}{24}\left(1+\frac{1}{n} \right)\Big(\log\left[\sinh\left(\frac{\pi(u-u')}{\beta}\right) \right]-\log\left[\sinh\left(\frac{\pi(u'+v)}{\beta}\right) \right]\nn \\
&&\phantom{=\frac{c}{24}\left(1+\frac{1}{n} \right)}-\log\left[\sinh\left(\frac{\pi(u+v')}{\beta}\right) \right]+\log\left[\sinh\left(\frac{\pi(v-v')}{\beta}\right) \right]\Big).
\eea
\end{widetext}
The summation of the above terms is equal to the timelike R\'enyi entropy. The imaginary part comes from the term (\ref{thermalimaginary}).

\subsection{ States dual to pure AdS$_3$}\label{holographicstate}

The general AdS$_3$ spacetime satisfying vacuum Einstein equation is given by the Banado geometry. By a conformal transformation $\xi=f(u)$, $\bar \xi=g(v)$, the geometry can be mapped to the Poincare patch. Using the conformal mapping one could evaluate the holographic entanglement entropy of an interval between $(t,x)$ and $(t',x')$ by RT formula. After analytic continuation the result is
\bea\label{holographicEE}
\hspace{-1em}S(t,x;t',x')=\frac{c}{6}\log \left[\frac{(f(u')-f(u-i\epsilon))(g(v-i\epsilon)-g(v'))}{\delta^2 \sqrt{f'(u-i\epsilon)f'(u')g'(v-i\epsilon)g'(v')} }\right],\nn
\eea
where $\epsilon>0$, $\delta$ is the UV cut-off. We would expect the conformal transformation to preserve causality. If $(t,x)$ and $(t',x')$ are timelike, we require that their images satisfy $(f(u)-f(u'))(g(v)-g(v'))<0$. The timelike entanglement entropy is given by
\bea
S(t,x;t',x')=\frac{c}{6}\log \left[\frac{(f(u)-f(u'))(g(v)-g(v'))}{\delta^2 \sqrt{f'(u)f'(u')g'(v)g'(v')} }\right]+\frac{i\pi c}{6}.\nn\\
~
\eea
In Appendix.\ref{app_B} we check the formula (\ref{linearcombinationalt}) hold for the state dual to AdS$_3$.

\subsection{Imaginary part of timelike entanglement entropy}
To understand the imaginary part of timelike EE, we can introduce the operator
\bea\label{comm_derivative}
\dot{\sigma}_n:= i[H^{(n)},\sigma_n],
\eea
where $H^{(n)}$ is the Hamiltonian of CFT$_n$. It can be shown that $\dot{\sigma}_n$ can be taken as time derivative with respect to Lorentzian time $t$.
Similarly, we define $\dot{\tilde{\sigma}}_n:=i[H,\tilde{\sigma}_n]$.  Therefore, we would have the correlators  involving of $\dot{\sigma}_n$ and $\dot{\tilde{\sigma}}_n$, such as $
\langle \Psi| \dot{\sigma}_n\tilde{\sigma}_n|\Psi\rangle$,  $\langle \Psi| \sigma_n\dot{\tilde{\sigma}}_n|\Psi\rangle$ and  $\langle \Psi| \dot{\sigma}_n\dot{\tilde{\sigma}}_n|\Psi\rangle$.
By the definition of R\'enyi entropy, we have
\bea
\p_{t}S_n=\frac{1}{1-n}\frac{\langle \Psi| \dot{\sigma}_n(t,x)\tilde{\sigma}_n(t',x')|\Psi\rangle}{\langle \Psi| \sigma_n(t,x)\tilde{\sigma}_n(t',x')|\Psi\rangle},
\eea 
or equally,
\bea
\langle \Psi| \dot{\sigma}_n(t,x)\tilde{\sigma}_n(t',x')|\Psi\rangle=\partial_t e^{(1-n)S_n}.
\eea
Applying Eq. (\ref{formrelation}) to the twist operator $\sigma_n$, we generally anticipate that operators from the given theory may appear in the integral on the right-hand side of (\ref{formrelation}). However, in all the states discussed above, it is noteworthy that other contributions vanish, except for the four operators $\sigma_n(0,x)$, $\tilde{\sigma}_n(0,x)$, $\dot{\sigma}_n(0,x)$, and $\dot{\tilde{\sigma}}_n(0,x)$.

All instances indicate that the imaginary part of timelike entanglement entropy originates from the term
\bea
\frac{1}{4}\int_{-u}^{v} d\bar x \partial_{t}S(0,\bar x;0,-u')+\frac{1}{4}\int_{-u}^{v} d\bar x \partial_t S(0,\bar x;0,v'),\nn
\eea
assuming $-u<-u'<v'<v$.
For the timelike EE, we have
\bea\label{imaginary_twist_derivative}
&&2 \text{Im}S(t,x;t',x')\nn \\
&&=\frac{1}{4}\int_{-u}^{v} d\bar x \lim_{n\to 1}\frac{1}{1-n} \langle \Psi| [\dot{\sigma}_n(0,\bar x),\tilde{\sigma}_n(0,-u')]|\Psi\rangle\nn \\
&&+\frac{1}{4}\int_{-u}^{v} d\bar x \lim_{n\to 1}\frac{1}{1-n} \langle \Psi| [\dot{\sigma}_n(0,\bar x),\tilde{\sigma}_n(0,v)]|\Psi\rangle.
\eea
Refer to supplemental material for additional details. The commutator $[\dot{\sigma}_n(0,\bar x),\tilde{\sigma}_n(0,\bar y)]=0$ for $\bar x\ne \bar y$ due to spacelike separation, yielding a delta function $\delta(\bar x-\bar y)$. For all considered states, the imaginary part of timelike EE remains a constant $\frac{i\pi c}{6}$. The expected commutator is
\bea
&&\langle \Psi|[\dot{\sigma}_n(0,\bar x),\tilde{\sigma}_n(0,\bar y)]|\Psi\rangle\nn \\
&&=\frac{2i\pi c}{3}(1-n)\delta(\bar x-\bar y)+O(1-n)^2.
\eea
Note that this commutator is specific to the discussed states; additional terms at order $O(1-n)$ may exist in more general cases, represented by $\mA_i(\bar x)\delta(\bar x-\bar y)$, satisfying $\langle \Psi| \mA_i |\Psi\rangle=0$.

Lastly, the time derivative of EE can be articulated in terms of entanglement spectra. By definition,
\bea
e^{(1-n)S_n}=tr \rho_A^n=\sum_i \lambda_i^n,
\eea
where $\lambda_i$ are eigenvalues of $\rho_A$.
Thus we have
\bea
\p_{t}  S=-\sum_i \p_t \lambda_i \log \lambda_i =- \int_0^{\lambda_m}d\lambda \log\lambda \ \mathcal{P}_t(\lambda), 
\eea
where $\sum_i \p_t\lambda_i=0$, and $\mathcal{P}_t(\lambda):=\sum_i \frac{\p \lambda_i}{\p t} \delta(\lambda_i-\lambda)$. The maximal eigenvalue is denoted by $\lambda_m$. The relationship between $\mathcal{P}_t(\lambda)$ and the density of eigenstates $\mathcal{P}(\lambda):=\sum_i \delta(\lambda_i-\lambda)$ is given by $\frac{\p \mathcal{P}}{\p t}=-\frac{\p \mathcal{P}_t}{\p \lambda}$ \cite{Guo:2023tys}.


\subsection{The Significance of the relation}



A significant observation is that terms without time derivatives in Eq. (\ref{linearcombination}) exclusively involve entanglement between four points at $t=0$, as illustrated in Fig. \ref{CausalDomain}. Scaling the UV cutoff $\delta \to c \delta$ results in a constant shift in both timelike and spacelike RE or EE. The constant from the left-hand side and the right-hand side cancel each other out thanks to the coefficients $\frac{1}{4}$. 

\begin{figure}
\centering
\subfigure[]{\includegraphics[scale=0.5]{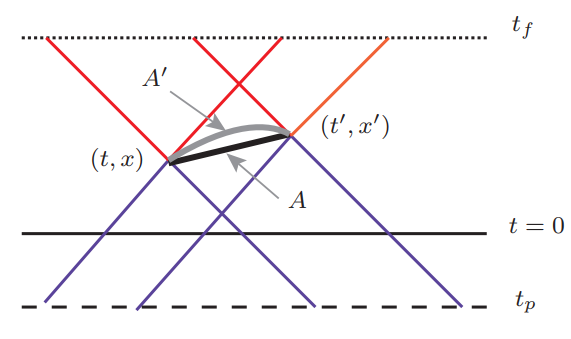}}
\subfigure[]{\includegraphics[scale=0.5]{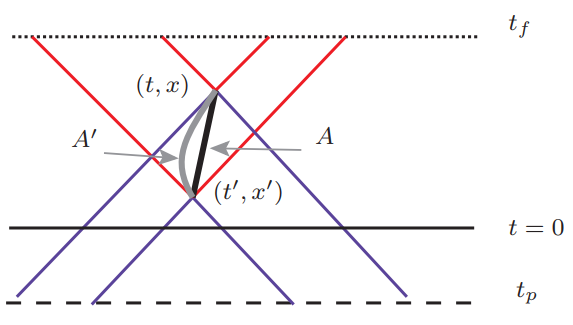}}
\caption{Spacelike and timelike seperations of $(t,x)$ and $(t',x')$. $t_p$ and $t_f$ are two time slices. (a) $(t,x)$ and $(t',x')$ are spacelike.   $A$ (black) and $A'$ (gray) has same causal domain of dependence, which is given by non-overlaping region between past and future lightcones of the two points. (b) $(t,x)$ and $(t',x')$ are timelike. $A$ (black) and $A'$ (gray) has same timelike envelope, which is the overlaping region between past lightcone of $(t,x)$ and future lightcone of $(t',x')$. }
\label{twocases}
\end{figure}

In the above derivation, we consider the Cauchy surface to be $t=0$. It is straightforward to generalize the result to other Cauchy surfaces, such as $t=t_p$ or $t=t_f$, as illustrated in Fig.\ref{twocases}. It is also worth noting that the relation (\ref{linearcombination}) remains valid when $(t,x)$ and $(t',x')$ are spacelike, as depicted in Fig.\ref{twocases}(a). Hence, timelike and spacelike EE can be unified within a single framework. For the spacelike case, it is easy to demonstrate that the right-hand side of (\ref{linearcombination}) will be real.

The relation (\ref{linearcombination}) also demonstrates that the timelike EE is solely linked to the causal region associated with the endpoints of the timelike interval, as depicted in Fig.\ref{CausalDomain} and Fig.\ref{twocases}. This suggests that the timelike EE follows the causal constraints and is intricately connected to the dynamics of the underlying theory.

In the spacelike case (Fig.\ref{twocases}(a)), $A$ and $A'$ have the same endpoints and causal domain of dependence $\mathcal{D}(A)=\mathcal{D}(A')$. Consequently, $A$ and $A'$ would possess the same entropy, consistent with the fact that the right-hand side of (\ref{linearcombination}) depends only on the endpoints $(t,x)$ and $(t',x')$. The reduced density matrix $\rho_A$ is associated with $\mathcal{D}(A)$, and the unitary operator $U(s):=\rho_A^{is}$ would transform the operators in $\mathcal{D}(A)$ into itself \cite{Casini:2011kv}.

\begin{figure}[htbp]
  \centering
  \includegraphics[width=0.2\textwidth]{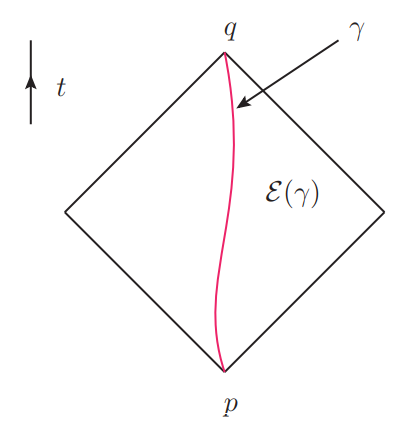}
  \caption{The timelike envelope $\mathcal{E}(\gamma)$ of the timelike wordline $\gamma$ (red curve), $p,q$ are end points of $\gamma$. In the simple case that we are interested in this paper, $\mathcal{E}$ is given by the intersection of the past of $q$ and future of $p$. }
  \label{envelope}
\end{figure}

 While for the timelike worldline $\gamma$, one can establish the algebra of operators within the timelike envelope $\mathcal{E}(\gamma)$. This envelope is defined by the set reachable through the deformation of $\gamma$ by timelike curves while maintaining the fixed endpoints of $\gamma$, as depicted in Fig. \ref{envelope}. According to the timelike tube theorem, the algebra of operators on the worldline $\gamma$ is equivalent to those within its timelike envelope $\mathcal{E}(\gamma)$ \cite{Borchers,Araki,Strohmaier:2023hhy,Strohmaier:2023opz}. Recent studies have demonstrated the utility of the algebra associated with an observer to comprehend the concept of entropy, particularly when gravity is taken into account\cite{Chandrasekaran:2022cip,Chandrasekaran:2022eqq,Witten:2023qsv,Witten:2023xze}. 

The relation (\ref{linearcombination}) implies that timelike EE can be understood in a manner analogous to spacelike EE. By a similar argument, we can show that the timelike EE for $A$ and $A'$ in Fig.~\ref{twocases}(b) is equal. Here $A'$ represents arbitrary curves with fixed endpoints $(t,x)$ and $(t',x')$. In fact, these curves form the set $\mathcal{E}(A)$, namely the timelike envelope of the interval $A$. The relation (\ref{linearcombination}) thus suggests that the timelike EE depends only on the endpoints of the subsystem. This is consistent with the recent explanation of timelike EE in \cite{Milekhin:2025ycm}.

The timelike EE associated with curves ending at $(t,x)$ and $(t',x')$ can be understood as the entanglement for the generalized density matrix defined by the subsystems $A_1 := (x,+\infty)$ at timeslice $t$ and $A_2 := (-\infty,x')$ at timeslice $t'$. Therefore, all curves in the timelike envelope correspond to the same generalized density matrix for $A_1$ and $A_2$, and consequently yield the same timelike EE.

Although we do not expect the timelike EE to be directly associated with the operator algebra of $\mathcal{E}(A)$, there should exist an operator $\mathcal{T}_{A_1A_2}$ determined by operators localized on $A_1$ and $A_2$. Moreover, the fact that different curves in the envelope $\mathcal{E}(A)$ give the same EE suggests the presence of an underlying symmetry transformation relating them. Understanding timelike EE from the perspective of operator algebras would be an interesting direction for future study. In particular, modular operators and modular flows in spacelike subsystems may admit analogous structures in timelike subsystems.

\section{General case }\label{General_section}
In the previous sections, we have focused on examples in CFT$_2$. In two dimensions, the notion of timelike EE can be understood more clearly, since the number of time and spatial coordinates is the same. This allows one to interpret the timelike EE by exchanging the roles of time and space. In such cases, the timelike EE can be computed via the analytic continuation of twist operator correlators. For more general states, the operator product expansion (OPE) can also be applied to evaluate the EE.

In higher-dimensional theories, however, additional subtleties arise. A more general definition of timelike EE can be formulated through the spacetime (or generalized) density matrix proposed in \cite{Milekhin:2025ycm, Gong:2025pnu}. Yet, it remains unclear how to explicitly evaluate the timelike EE in both QFTs and holography. Several holographic proposals have been made \cite{Doi:2022iyj, Heller:2024whi}. It is an interesting open question whether the relation established in this paper can be extended to more general situations, including generic states in CFT$_2$ and higher-dimensional cases. We will briefly comment on this point in the following.

\subsection{Timelike EE via OPE of twist operators}

However, it is crucial to emphasize that the relation (\ref{linearcombination}) does not hold universally. This can be demonstrated by delving into the operator product expansion (OPE) of twist operators—a valuable tool for computing RE in QFTs \cite{Cardy:2007mb,Headrick:2010zt,Calabrese:2010he,Rajabpour:2011pt,Chen:2013kpa}. The OPE of twist operators in CFT$_n$ can be formally written as
\bea
&&\hspace{-1em}\sigma_n(w,\bar w)\tilde{\sigma}_n(w',\bar w')\nn \\
&&\hspace{-1em}=\sum_K d_K (w-w')^{-2h_n+h_K}(\bar w-\bar w')^{-2\bar h_n+\bar h_K}\mathcal{X}_{K},
\eea 
where $\mathcal{X}_{K}$ denotes the operators in CFT$_n$, and $d_K$ are the OPE coefficients. The computation of RE and EE involves evaluating the expectation value $\langle \Psi|\mathcal{X}_K|\Psi\rangle$ and determining the coefficients $d_K$. To explore EE for arbitrary intervals, analytical continuation is employed, yielding a result expressed as an infinite summation:
\bea\label{EEshort}
&&\hspace{-1em}S(t,x;t',x')\nn \\
&&\hspace{-1em}=S_0(t,x;t',x')+\sum_k a_k  [-(u-u'-i\epsilon)]^{2h_k}[v-v'-i\epsilon]^{2\bar h_k},\nn\\
~
\eea
where $S_0$ denotes the EE for the vacuum state.
Note that we include the OPE coefficients and the expectation values of the operators in the coefficients $a_k$. For simplicity, in the following we will consider $a_k$ as constants, which is reasonable when the length of the interval is short enough.

Since $S_0$ satisfies the relation (\ref{linearcombination}), we only need to consider the summation terms in (\ref{EEshort}). In the following, we still assume $-u<-u'<v'<v$. For the timelike separation $-u+u'<0$, there may be new imaginary contributions to the timelike EE from the summation terms. One could directly check that generally, the result (\ref{EEshort}) \textit{does not} satisfy the relation (\ref{linearcombination}). 

We will focus in this section on a specific class of states where the OPE of twist operators includes only contributions from the vacuum family. These results will help us understand why the relation (\ref{linearcombination}) remains valid for thermal states and for states with holographic duals in AdS$_3$.

If the summation terms are the following form
\bea\label{factor}
\sum_m a_m \left([-(u-u'-i\epsilon)]^{m}+[v-v'-i\epsilon]^{m}\right),
\eea
where $m$ are positive integers, we would like to show the relation (\ref{linearcombination}) is satisfied in the above case. 
In general the OPE of twist operators cannot be factored as a product of holomorphic and anti-holomorphic parts\cite{Cardy:2007mb,Headrick:2010zt,Calabrese:2010he,Rajabpour:2011pt,Chen:2013kpa}. This is because the operators $\mathcal{X}_K(w,\bar w)\ne \mathcal{X}_K(w)\mathcal{X}_{ K}(\bar w)$. However, if we only consider contributions from the vacuum family, the operators would be descendants of the identity operators, such as the stress-energy tensor $T(w)$ and $\bar{T}(\bar w)$. The OPE of $T$ and $\bar{T}$ is trivial, thus the contributions from $T(w)\bar{T}(\bar w)$ can be written as the product of the holomorphic and anti-holomorphic parts $\langle T(w)\rangle_\psi \langle \bar{T}(\bar w)\rangle_\psi$. It is also important that the OPE coefficients $d_K$ for the operator $T(w)\bar{T}(\bar w)$ also factor as $d_{T\bar T}=d_T d_{\bar T}$. The argument can be generalized to other higher-dimensional descendants. As a result, the OPE of twist operators can be written as $\log\langle \sigma_n(w,\bar w)\tilde{\sigma}_n(w',\bar w')$ can be expressed as a summation over the holomorphic and anti-holomorphic parts. Thus, the EE takes the form as in Eq.(\ref{factor}).

Now let us check if the form Eq.(\ref{factor}) does satisfy the relation Eq.(\ref{linearcombination}).
We only need to consider the terms $\Delta S:= S(t,x;t',x')-S_0(t,x;t',x')$.
 This can be done by directly computing. We have
 \begin{widetext}
\bea\label{g1}
&&\frac{1}{4}\left(\Delta S(0,-u;0,-u')+\Delta S(0,-u;0,v')+\Delta S(0,v;0,-u')+\Delta S(0,v;0,v') \right)\nn\\
&&=\sum_m a_m\Big[\left(u'+v+i \epsilon \right)^m+\left(u'+v-i \epsilon \right)^m+\left(u'-u+i \epsilon \right)^m+\left(u'-u-i \epsilon \right)^m\nn \\
&&+\left(-u-v'+i \epsilon \right)^m+\left(-u-v'-i \epsilon \right)^m+\left(-v'+v+i \epsilon \right)^m+\left(-v'+v-i \epsilon \right)^m\Big].\nn\\
\eea
\end{widetext}
It can also be shown there is no contribution from the following terms,
\bea
&&\frac{1}{4}\int_{-u'}^{v'} d\bar x'\partial_{t'}\Delta S(0,-u;0,\bar x')+\frac{1}{4}\int_{-u'}^{v'} d\bar x'\partial_{t'}\Delta S(0,v;0,\bar x')\rangle\nn \\
&&+\frac{1}{4}\int_{-u}^{v} d\bar x \partial_{t}\Delta S(0,\bar x;0,-u')+\frac{1}{4}\int_{-u}^{v} d\bar x \partial_t \Delta S(0,\bar x;0,v').\nn \\
~
\eea
By using 
\bea
&&\partial_t\partial_{t'}\Delta S(0,\bar x;0,\bar x')\nn \\
&&=\sum_m  a_m\left[(m-1) m \left(-(x-\bar x+i \epsilon )^{m-2}-(x-\bar x-i \epsilon )^{m-2}\right)\right],\nn
\eea
we can obtain
\bea\label{g2}
\frac{1}{4}\int_{-u}^{v} d\bar x \int_{-u'}^{v'} d\bar x' \partial_t\partial_{t'}\Delta S(0,\bar x;0,\bar x').
\eea
It is easy to check that the sum of (\ref{g1}) and (\ref{g2}) is given by
\bea
\sum_m a_m \left[ (-1)^m(u-u')^m+(v-v')^m\right],
\eea
which is just the $\Delta S$ for the timelike EE. As a check of the above calculations, one could expand the thermal state in the low-temperature limit. In this case only even $m$ contributes, thus the result is $\sum_m a_m \left[ (u-u')^m+(v-v')^m\right]$. By taking the coefficients $a_m$ for the thermal states one could obtain the timelike EE for thermal states. It can be shown that the thermal states or the general states dual to pure AdS$_3$ can be written in the above form. 

Let us explain why the above form is correct for the states dual to AdS$_3$. It is known that these theories have a sparse light spectrum \cite{Hartman:2014oaa}. The main contributions to the OPE of twist operators come from the vacuum family, such as the lowest contributions from the stress-energy tensor $T$ and $\bar{T}$ \cite{Asplund:2014coa}. Thus, the powers of the expansion (\ref{factor}) should be integers, and $m\geq 2$. It has been shown that in this case, the OPE can be factored into the product of holomorphic and anti-holomorphic parts. Consequently, the EE can be expanded in the form (\ref{factor}). This can also be taken as a proof of the relation (\ref{linearcombination}) for thermal states and states with holographic duals using a perturbative approach.


\subsection{Higher dimensional examples}

In higher-dimensional examples, one can evaluate the holographic timelike EE through the RT formula. In \cite{Doi:2022iyj}, the holographic timelike EE was interpreted in terms of spacelike and timelike geodesic lines in AdS$_3$. In contrast, \cite{Heller:2024whi} proposed that the dual RT surfaces should be extended into a complexified geometry. In AdS$_3$, both proposals lead to the same result. However, in higher-dimensional cases, they generally yield different outcomes, as also reflected in the distinct approaches developed in \cite{Guo:2025pru, Nunez:2025ppd, Heller:2025kvp} for computing holographic timelike EE.

In \cite{Heller:2024whi, Nunez:2025ppd, Heller:2025kvp}, the authors computed the holographic EE for strip subregions that can have either spacelike or timelike separations. They proposed that the timelike EE can be obtained via an analytic continuation of a parameter which explores both timelike and spacelike cases. In their framework, the timelike and spacelike EEs are connected by analytic continuation of suitable parameters.

In this work, we establish a relation between timelike and spacelike EEs, but of a different nature from that in \cite{Heller:2024whi, Nunez:2025ppd, Heller:2025kvp}. Specifically, the relation (\ref{linearcombination}) connects the EEs of causally separated (timelike) and spacelike subregions, as discussed in Section~\ref{TEE_SEE_relation_section}. This relation arises naturally from causality constraints, rather than from analytic continuation. To establish the relation, one must first know the explicit form of the timelike EE. In principle, one could attempt to construct a similar relation as (\ref{linearcombination}) in the examples considered in \cite{Heller:2024whi, Nunez:2025ppd, Heller:2025kvp}.

It remains an interesting question whether the relation (\ref{linearcombination}) can be generalized to higher-dimensional theories. Partial evidence was found in \cite{Guo:2024ITE}, where the imaginary part of the timelike EE for timelike-separated strip subregions in the higher-dimensional vacuum was shown to be related to the spacelike EE. Furthermore, in \cite{Guo:2025pru}, the timelike EE in higher-dimensional setups was proposed to be computed via analytic continuation from Euclidean results, consistent with the methods developed in \cite{Heller:2024whi, Nunez:2025ppd, Heller:2025kvp}. In specific examples, a similar relation indeed emerges \cite{Guo:2025pru}. However, for more general subregions and arbitrary states, whether such a correspondence universally holds remains an open question.

\section{Conclusion and Discussion}\label{discuss_section}
In this paper, we establish a connection between timelike and spacelike EE. Our results show that, within a broad class of states, timelike EE can be expressed as a linear combination of spacelike EE and its first derivative. In 2D CFTs, we demonstrate that states conformally equivalent to the vacuum satisfy this relation. For more general states, the relation (\ref{linearcombination}) requires modifications, but using the OPE of twist operators we show that it can be constructed perturbatively. These findings highlight that timelike and spacelike EE can be unified within a single framework, suggesting that timelike EE naturally captures the entanglement structure of spacetime subsystems. This indicates that the notion of timelike EE is well-motivated in QFTs and is deeply tied to causality and dynamics.

Our discussion here primarily focuses on field theory, where the sum rule for timelike EE emerges as a consequence of causality. Remarkably, the timelike correlator of the twist operator exhibits such a simple relation to its spacelike counterpart, even for states beyond the vacuum. In particular, we find that the relation (\ref{linearcombination}) holds for holographic states. This raises the intriguing question of its gravitational dual: timelike EE is expected to correspond to a bulk RT surface \cite{Doi:2022iyj}, suggesting a direct connection between timelike and spacelike RT surfaces. Thus, the relation (\ref{linearcombination}) can be viewed as a manifestation of intrinsic geometric structures in the bulk. Recently, \cite{Guo:2025pru} proposed that this relation may be understood as a duality between RT surfaces inside and outside the black hole horizon, thereby providing an exact correspondence between interior and exterior degrees of freedom. This further implies that one may probe the black hole interior solely through exterior data. Consequently, the relation between timelike and spacelike EE carries profound physical significance, especially in the context of black hole physics and holography. It would be interesting to extend this relation to more general backgrounds and to explore its geometric interpretation in greater detail.

Applying the RT formula to the dS spacetime, it can be shown that the holographic EE is also complex \cite{Doi:2022iyj}\cite{Narayan:2022afv}. In the context of the dS/CFT correspondence \cite{Strominger:2001pn}, the result suggests it should be explained as pseudoentropy in the dual CFTs \cite{Doi:2022iyj}. A notable fact is that the expression of holographic EE in dS/CFT is very similar to the timelike EE in the vacuum state. It would be interesting to explore in the near future whether there exists a similar relation as (\ref{linearcombination}) for the holographic EE in dS/CFT. 

An additional noteworthy observation is that timelike entanglement entropy finds an interpretation through pseudoentropy. As demonstrated in \cite{Guo:2023aio}, a sum rule links pseudo-R\'enyi entropy to R\'enyi entropy. An intriguing avenue for future exploration is to investigate whether the relation (\ref{linearcombination}) serves as a specific instance of the general sum rule proposed in \cite{Guo:2023aio}.

{\bf Acknowledgements}
We would like to thank Yan Liu, Rong-Xin Miao, Tadashi Takayanagi, Run-Qiu Yang, Jiaju Zhang for the useful discussions.
WZG is supposed by the National Natural Science Foundation of China under Grant No.12005070 and the Hubei Provincial Natural Science Foundation of China under Grant No.2025AFB557. SH would like to appreciate the financial support from the Fundamental Research Funds for the Central Universities, Max Planck Partner Group, and the Natural Science Foundation of China Grants (No. 12475053 and No. 12235016).

\appendix
\begin{widetext}
\section{Canonical quantization of two-dimensional massless free theory}\label{app_A}

In this section, we initially present two methods for deducing Eq.(\ref{operatorsumrule}).
This equation is crucial to derive the timelike EE relation in the vacuum state.

\paragraph{Method 1:} 
The Euclidean action of a massless free scalar theory is given by
\bea
S=\frac{1}{4\pi \kappa}\int dz d\bar z \p_z \phi \p_{\bar z}\phi, 
\eea
where $z=\tau+i x$ and $\bar z=\tau- ix$.
We can obtain the two-point correlators
\bea
&&\langle  \phi(z,\bar z) \phi(z',\bar z')\rangle =-\kappa\log \left[(z-z')(\bar z-\bar z')\right]\nn \\
&&\phantom{\langle  \phi(z,\bar z) \phi(z',\bar z')\rangle}=-\kappa \log\left[(\tau-\tau')^2+(x-x')^2 \right], 
\eea
where $z=\tau+i x$ and $z'=\tau'+i x'$. We can use the $i\epsilon$ prescription to obtain the correlator in Minkowski spacetime through analytical continuation $\tau\to i t+\epsilon$ and $\tau'\to i t+\epsilon'$, e.g., with $\epsilon> \epsilon'$ we get
\bea\label{phiphicorrelator}
\langle  \phi(t,x) \phi(t',x')\rangle=-\kappa \log\left[ \Delta s^2+2i(\epsilon-\epsilon')(t-t') \right],
\eea
where $\Delta s^2=-(t-t')^2+(x-x')^2$.

By canonical quantization we can expand the operators $\phi(t,x)$ and $\pi(t,x)$ as
\bea\label{canonicalquantization}
&&\phi(t,x)=\int_{-\infty}^{\infty}\frac{dk}{2\pi} \frac{1}{\sqrt{2 w_k}}\left(a_k e^{-i w_k t+i kx}+a_k^\dagger e^{i w_k t-i k x} \right),\nn \\
&&\pi(t,x)=\frac{1}{4\pi \kappa}\int_{-\infty}^{\infty}\frac{dk}{2\pi} \frac{-iw_k}{\sqrt{2 w_k}}\left(a_k e^{-i w_k t+i kx}-a_k^\dagger e^{i w_k t-i k x} \right),
\eea
where $w_k=|k|$ and $[a_k,a_{k'}^\dagger]=8\pi^2 \kappa \delta(k-k')$. We can obtain the correlator 
\bea\label{correlatorpp}
&&\langle \phi(t,x)\phi(t',x')\rangle=4\pi \kappa \int_0^{+\infty}\frac{dk}{2\pi} \frac{1}{2k} e^{-i k (u-u')}+4\pi \kappa \int_0^{+\infty}\frac{dk}{2\pi} \frac{1}{2k} e^{-i k (v-v')},
\eea
where $u:=t-x$, $v:=t+x$. By using $\int_0^{+\infty}dk e^{ikx-k \epsilon}=\frac{i}{x+i \epsilon}$ for $\epsilon>0$ we can obtain the correlator $\langle  \phi(t,x) \phi(t',x')\rangle$ (\ref{phiphicorrelator}). Further, we can obtain 
\bea\label{correlatorphipi_canonical}
&&\langle \phi(t,x)\pi(t',x')\rangle= \frac{i}{2} \int_0^{+\infty}\frac{dk}{2\pi}e^{-i k(u-u')}+\frac{i}{2}\int_0^{+\infty}\frac{dk}{2\pi}e^{-i k (v-v')}\nn \\
&&\phantom{\langle \phi(t,x)\pi(t',x')\rangle}=\frac{1}{4\pi}\frac{1}{u-u'- i\epsilon}+\frac{1}{4\pi}\frac{1}{v-v'-i\epsilon},
\eea
which can be related to (\ref{phiphicorrelator}) by 
\bea
&&\langle \phi(t,x)\pi(t',x')\rangle=\frac{1}{4\pi \kappa}(\p_{v'}+\p_{u'})\langle \phi(t,x)\phi(t',x')\rangle\nn \\
&&\phantom{\langle \phi(t,x)\pi(t',x')\rangle}=-\frac{1}{4\pi \kappa}(\p_v+\p_u)\langle \phi(t,x)\phi(t',x')\rangle.
\eea
We can further obtain
\bea
\langle \pi(t,x)\pi(t',x')\rangle=\frac{1}{(4\pi \kappa)^2}(\p_{v'}+\p_{u'})(\p_{v}+\p_{u})\langle \phi(t,x)\phi(t',x')\rangle.\nn 
\eea
It is possible to compute $f_{\phi}$ and $f_{\pi}$ by using (\ref{canonicalquantization}). Since $f_{\phi}$ and $f_{\pi}$ are numbers, we have 
\bea
f_{\pi}(t,x;t',x')=\langle f_{\pi}(t,x;t',x')\rangle=i\langle \phi(t,x)\phi(t',x')\rangle-i\langle \phi(t',x') \phi(t,x)\rangle,\nn \\
f_{\phi}(t,x;t',x')=\langle f_{\phi}(t,x;t',x')\rangle=-i\langle \phi(t,x)\pi(t',x')\rangle+i\langle \pi(t',x') \phi(t,x)\rangle, \nn
\eea
 which can be evaluated by using the correlators (\ref{phiphicorrelator}) and (\ref{correlatorphipi_canonical}). The results are
\bea\label{ffphipi_app}
&&f_{\pi}(t,x;t',x')
=2\pi\kappa\left(H(u-u')-H(v'-v)\right) ,\nn \\
&& f_{\phi}(t,x;t',x')=\frac{1}{2} \left(\delta(u-u')+\delta(v-v') \right).
\eea
$f_\phi$ is associated with  $f_\pi$ by the relation
\bea
 &&f_{\phi}(t,x;t',x')=-\frac{1}{4\pi \kappa}(\p_{v'}+\p_{u'})f_{\pi}(t,x;t',x')\nn \\
 &&\phantom{f_{\phi}(t,x;t',x')}=\frac{1}{4\pi \kappa}(\p_v+\p_u)f_{\pi}(t,x;t',x').
\eea
\paragraph{Method 2:} In the second method, we first leverage the Fourier transformation to find the inverse expression of $a_k$ and $a^\dagger_k$ with respect to $\phi(0,x)$ and $\pi(0,x)$,
\begin{align}
a_{k}=&\int_{-\infty}^{+\infty} d x\left(\sqrt{\frac{w_{ k}}{2}}\phi(0,x)+\frac{4\pi i\kappa}{\sqrt{2w_{k}}}\pi(0,x)\right)e^{-ikx},\nn\\
a^\dagger_{k}=&\int^{+\infty}_{-\infty} d x\left(\sqrt{\frac{w_{ k}}{2}}\phi(0, x)-\frac{4\pi i\kappa}{\sqrt{2w_{ k}}}\pi(0,x)\right)e^{i kx}.\label{eq15a}
\end{align}
Substituting Eq. \eqref{eq15a} into Eq. \eqref{canonicalquantization}, we then obtain the expression of $\phi(t,x)$ in terms of $\phi(0,x)$ and $\pi(0,x)$, 
\begin{align}\label{fjfj}
\phi(t,x)=\int^{+\infty}_{-\infty} d\bar x \Big(f_{\phi}(t,x;0,\bar x)\cdot\phi(0,\bar x)+f_{\pi}(t,x;0,\bar x)\cdot\pi(0,\bar x)\Big),
\end{align}
where
\begin{align}
f_{\phi}(t,x;0,\bar x)\equiv &\frac{1}{2}\int^{+\infty}_{-\infty}\frac{dk}{2\pi}\left(e^{-iw_kt+ik(x-\bar x )}+e^{iw_kt-ik(x-\bar x)}\right),\nonumber\\
f_{\pi}(t,x;0,\bar x)\equiv&2\pi\kappa\int_{-\infty}^{+\infty}\frac{dk}{2\pi}\left(\frac{i}{w_k}e^{-iw_kt+ik(x-\bar x)}-\frac{i}{w_k}e^{iw_kt-ik ( x-\bar x)}\right).\label{eq25}
\end{align}
The first integral in \eqref{eq25} is easy to figure out since we are attending to a massless free scalar ($w_k=|k|$). It turns out that 
\begin{align}
f_\phi(t,x;0,\bar x)=\frac{1}{2}\Big(\delta(x-t-\bar x)+\delta(x+t-\bar x)\Big).\label{hhf}
\end{align}
On the other hand, by comparing two integrals in \eqref{eq25}, we can find a relation between $f_\phi$ and $f_\pi$ 
\begin{align}
f_{\phi}(t,x;0,\bar x)=\frac{1}{4\pi\kappa}\partial_tf_{\pi}(t,x;0,\bar x). \label{gh12}
\end{align}
Combining the above relation with Eq. \eqref{hhf}, we can write down the expression of $f_\pi$ immediately, 
\begin{align}\label{fasf}
f_{\pi}(t,x;0,\bar x)=2\pi\kappa\Big(H(x+t-\bar x)-H(x-t-\bar x)\Big)=
\begin{cases}
2\pi\kappa,\quad \bar x\in[x-t,x+t],\\
0,\quad \text{otherwise}.
\end{cases}
\end{align}
Note that $f_\phi$ in Eq. \eqref{gh12} and $f_\pi$ in Eq. \eqref{fasf}, up to a time translation, are equivalent to $f_\phi$ and $f_\pi$ in Eq. \eqref{ff_main}, respectively.

Substituting \eqref{gh12} and \eqref{fasf} into \eqref{fjfj}, we arrive at an expression
\bea\label{operatorsumrule}
\phi(t,x)=\frac{1}{2}\phi(0,-u)+\frac{1}{2}\phi(0,v)+2\pi \kappa \int_{-u}^{v}d\bar x \pi(0,\bar x).
\eea

One could directly check this result by using the canonical quantization formula (\ref{canonicalquantization}).
We can rewrite (\ref{canonicalquantization}) as 
\bea
&&\phi(t,x)=\int_0^{+\infty}\frac{dk}{2\pi}\frac{1}{\sqrt{2k}}\left(a_k e^{-ik(t-x)}+a_k^\dagger e^{ik(t-x)} \right)\nn \\
&&\phantom{\phi(t,x)=}+\int_0^{+\infty}\frac{dk}{2\pi}\frac{1}{\sqrt{2k}}\left(a_{-k} e^{-ik(t+x)}+a_{-k}^\dagger e^{ik(t+x)} \right), \nn \\
&&\pi(t,x)=\frac{1}{2\pi \kappa}\int_0^{+\infty}\frac{dk}{2\pi}\frac{-ik}{\sqrt{2k}}\left(a_k e^{-ik(t-x)}-a_k^\dagger e^{ik(t-x)} \right)\nn \\
&&\phantom{\pi(t,x)=}+\frac{1}{2\pi \kappa}\int_0^{+\infty}\frac{dk}{2\pi}\frac{-ik}{\sqrt{2k}}\left(a_k e^{-ik(t+x)}-a_k^\dagger e^{ik(t+x)} \right).
\eea
Thus we have
\bea
&&\phi(0,-u)=\int_0^{+\infty}\frac{dk}{2\pi}\frac{1}{\sqrt{2k}}\left(a_k e^{-iku}+a_k^\dagger e^{iku} \right)\nn \\
&&\phantom{\phi(0,-u)=}+\int_0^{+\infty}\frac{dk}{2\pi}\frac{1}{\sqrt{2k}}\left(a_{-k} e^{iku}+a_{-k}^\dagger e^{-iku} \right),\nn\\
&&\phi(0,v)=\int_0^{+\infty}\frac{dk}{2\pi}\frac{1}{\sqrt{2k}}\left(a_k e^{ikv}+a_k^\dagger e^{-ikv} \right)\nn \\
&&\phantom{\phi(0,v)=}+\int_0^{+\infty}\frac{dk}{2\pi}\frac{1}{\sqrt{2k}}\left(a_{-k} e^{-ikv}+a_{-k}^\dagger e^{ikv} \right),
\eea
and
\bea
&&\int_{-u}^v \pi(0,\bar x)d\bar x=\frac{1}{2\pi \kappa}\int_0^{+\infty}\frac{dk}{2\pi}\frac{-1}{\sqrt{2k}}\left( a_k e^{ikv}+a_k^\dagger e^{-ikv} \right)\nn \\
&&\phantom{\int_{-u}^v \pi(0,\bar x)d\bar x=}-\frac{1}{2\pi \kappa}\int_0^{+\infty}\frac{dk}{2\pi}\frac{-1}{\sqrt{2k}}\left( a_k e^{-iku}+a_k^\dagger e^{iku} \right)\nn \\
&&\phantom{\int_{-u}^v \pi(0,\bar x)d\bar x=}+\frac{1}{2\pi \kappa}\int_0^{+\infty}\frac{dk}{2\pi}\frac{1}{\sqrt{2k}}\left( a_{-k} e^{-ikv}+a_{-k}^\dagger e^{ikv} \right)\nn \\
&&\phantom{\int_{-u}^v \pi(0,\bar \pi)d\bar x=}-\frac{1}{2\pi \kappa}\int_0^{+\infty}\frac{dk}{2\pi}\frac{1}{\sqrt{2k}}\left( a_{-k} e^{iku}+a_{-k}^\dagger e^{-iku} \right)\nn \\
\eea
Using the above formulas, we can directly verify that Eq.(\ref{operatorsumrule}) is correct. 
\section{Calculation details for the pure AdS$_3$}\label{app_B}
In Section~\ref{holographicstate}, we discuss examples in the case of pure AdS$_3$. In the appendix, we verify that the relation (\ref{linearcombination}) holds for this example.
For the spacelike separation, we also have
\bea\label{general_ptp}
\partial_{t'}S(t,x;t',x')
=\frac{1}{12} c \left(-\frac{2f'\left(u'\right)}{f(u-i\epsilon)-f\left(u'\right)}-\frac{f''\left(u'\right)}{f'\left(u'\right)}-\frac{2g'\left(v'\right)}{g(v-i\epsilon)-g\left(v'\right)}-\frac{g''\left(v'\right)}{g'\left(v'\right)}\right),
\eea
\bea
\partial_{t}S(t,x;t',x')=\frac{1}{12} c \left(\frac{2f'(u)}{f(u-i\epsilon)-f\left(u'\right)}-\frac{f''(u)}{f'(u)}+\frac{2g'(v)}{g(v-i\epsilon)-g\left(v'\right)}-\frac{g''(v)}{g'(v)}\right),
\eea
and
\bea
\partial_{t}\partial_{t'}S(t,x;t',x')=\frac{1}{6} c \left(\frac{f'(u) f'\left(u'\right)}{\left(f(u-i \epsilon)-f\left(u'\right)\right)^2}+\frac{g'(v) g'\left(v'\right)}{\left(g(v-i \epsilon)-g\left(v'\right)\right)^2}\right).
\eea
Now we could use the above results to evaluate the right hand side of (\ref{linearcombination}). The first terms are
\bea
&&\frac{1}{4}\left(S(0,-u;0,-u')+S(0,-u;0,v')+S(0,v;0,-u')+S(0,v;0,v') \right)\nn \\
&&=\frac{c}{24}\Big(\log \left[\frac{(f(u)-f(u'))(g(-u)-g(-u'))}{\delta^2 \sqrt{f'(u)f'(u')g'(-u)g'(-u')}} \right]+\log \left[\frac{(f(u)-f(-v'))(g(-u)-g(v'))}{\delta^2 \sqrt{f'(u)f'(-v')g'(-u)g'(v')}} \right]\nn \\
&&\phantom{\frac{c}{24}\Big(}+\log \left[\frac{(f(-v)-f(u'))(g(v)-g(-u'))}{\delta^2 \sqrt{f'(-v)f'(u')g'(v)g'(-u')}} \right]+\log \left[\frac{(f(-v)-f(-v'))(g(v)-g(v'))}{\delta^2 \sqrt{f'(-v)f'(-v')g'(v)g'(v')}} \right]\Big).\nn\\
~
\eea
By using (\ref{general_ptp}) we have
\bea
&&\frac{1}{4}\int_{-u'}^{v'} d\bar x'\partial_{t'}S(0,-u;0,\bar x')\nn \\
&&=\frac{c}{48}\int_{-u'}^{v'} d\bar x'\left(-\frac{2f'\left(-\bar x'\right)}{f(u-i\epsilon)-f\left(-\bar x'\right)}-\frac{f''\left(-\bar x'\right)}{f'\left(-\bar x'\right)}-\frac{2g'\left(\bar x'\right)}{g(-u-i\epsilon)-g\left(\bar x'\right)}-\frac{g''\left(\bar x'\right)}{g'\left(\bar x'\right)}\right)\nn\\
&&=\frac{c}{48}\left(\log \frac{f'(-v')(g(-u)-g(v'))^2}{(f(u)-f(-v'))^2g'(v')}- \log \frac{f'(u')(g(-u)-g(-u'))^2}{(f(u)-f(u'))^2g'(-u')}\right),
\eea
and
\bea
&&\frac{1}{4}\int_{-u'}^{v'} d\bar x'\partial_{t'}S(0,v;0,\bar x')\nn \\
&&=\frac{c}{48}\int_{-u'}^{v'} d\bar x'\left(-\frac{2f'\left(-\bar x'\right)}{f(-v+i\epsilon)-f\left(-\bar x'\right)}-\frac{f''\left(-\bar x'\right)}{f'\left(-\bar x'\right)}-\frac{2g'\left(\bar x'\right)}{g(v-i\epsilon)-g\left(\bar x'\right)}-\frac{g''\left(\bar x'\right)}{g'\left(\bar x'\right)}\right)\nn\\
&&=\frac{c}{48}\left(\log \frac{f'(-v')(g(v)-g(v'))^2}{(f(-v)-f(-v'))^2g'(v')}- \log \frac{f'(u')(g(v)-g(-u'))^2}{(f(-v)-f(u'))^2g'(-u')}\right).
\eea
The next terms are
\bea
&&\frac{1}{4}\int_{-u}^{v} d\bar x \partial_{t}S(0,\bar x;0,-u')\nn \\
&&=\frac{c}{48}\int_{-u}^{v} d\bar x \left(\frac{2f'(-\bar x)}{f(-\bar x-i\epsilon)-f\left(u'\right)}-\frac{f''(-\bar x)}{f'(-\bar x)}+\frac{2g'(\bar x)}{g(\bar x-i\epsilon)-g\left(-u'\right)}-\frac{g''(\bar x)}{g'(\bar x)}\right)\nn\\
&&=\frac{c}{48}\int_{-u}^{v} d\bar x \left(\frac{2f'(-\bar x)}{f(-\bar x)-f\left(u'\right)-i \epsilon f'(-\bar x)}-\frac{f''(-\bar x)}{f'(-\bar x)}+\frac{2g'(\bar x)}{g(\bar x)-g\left(-u'\right)-i\epsilon g'(\bar x)}-\frac{g''(\bar x)}{g'(\bar x)}\right)\nn \\
&&=\frac{c}{48}\text{p.v.}\int_{-u}^{v} d\bar x \left(\frac{2f'(-\bar x)}{f(-\bar x)-f\left(u'\right)}-\frac{f''(-\bar x)}{f'(-\bar x)}+\frac{2g'(\bar x)}{g(\bar x)-g\left(-u'\right)+}-\frac{g''(\bar x)}{g'(\bar x)} \right)+\frac{ic\pi}{12}\nn\\
&&=\frac{c}{48}\log \frac{(f(u)-f(u'))^2f'(-v)}{(f(-v)-f(u'))^2f'(u)}+\frac{c}{48}\log \frac{(g(v)-g(-u'))^2g'(-u)}{(g(-u)-g(-u'))^2g'(v)}+\frac{ic\pi}{12}
\eea
and
\bea
&&\frac{1}{4}\int_{-u}^{v} d\bar x \partial_t S(0,\bar x;0,v')\nn \\
&&=\frac{c}{48}\int_{-u}^{v}d\bar x\left(\frac{2f'(-\bar x)}{f(-\bar x-i\epsilon)-f\left(-v'\right)}-\frac{f''(-\bar x)}{f'(-\bar x)}+\frac{2g'(\bar x)}{g(\bar x-i\epsilon)-g\left(v'\right)}-\frac{g''(\bar x)}{g'(\bar x)}\right)\nn \\
&&=\frac{c}{48}\log \frac{(f(u)-f(-v'))^2f'(-v)}{(f(-v)-f(-v'))^2f'(u)}+\frac{c}{48}\log \frac{(g(v)-g(v'))^2g'(-u)}{(g(-u)-g(v'))^2g'(v)}+\frac{ic\pi}{12}.
\eea
The last term is
\bea
&&\frac{1}{4}\int_{-u}^{v} d\bar x \int_{-u'}^{v'} d\bar x' \partial_t\partial_{t'}S(0,\bar x;0,\bar x')\nn \\
&&=\frac{c}{24}\int_{-u}^{v} d\bar x \int_{-u'}^{v'} d\bar x'   \left(\frac{f'(-\bar x) f'\left(-\bar x'\right)}{\left(f(-\bar x-i\epsilon)-f\left(-\bar x'\right)\right)^2}+\frac{g'(\bar x) g'\left(\bar x'\right)}{\left(g(\bar x-i\epsilon)-g\left(\bar x'\right)\right)^2}\right)\nn\\
&&=\frac{c}{24}\int_{-u}^{v} d\bar x \Big[-\frac{f'(-\bar x)}{f(-\bar x- i\epsilon)-f(-v')}+\frac{f'(-\bar x)}{f(-\bar x- i\epsilon)-f(u')}\nn \\
&&\phantom{=\frac{c}{24}\int_{-u}^{v} d\bar x \Big[}+\frac{g'(\bar x)}{g(\bar x- i\epsilon)-g(v')}-\frac{g'(\bar x)}{g(\bar x- i\epsilon)-g(-u')} \Big]\nn \\
&&=\frac{c}{24}p.v.\int_{-u}^{v} d\bar x \Big[-\frac{f'(-\bar x)}{f(-\bar x)-f(-v')}+\frac{f'(-\bar x)}{f(-\bar x)-f(u')}+\frac{g'(\bar x)}{g(\bar x)-g(v')}-\frac{g'(\bar x)}{g(\bar x)-g(-u')} \Big]\nn \\
&&=\frac{c}{24}\Big[-\log \frac{f(u)-f(-v')}{f(-v)-f(-v')}+\log \frac{f(u)-f(u')}{f(-v)-f(u')}+\log \frac{g(v)-g(v')}{g(-u)-g(v')}-\log \frac{g(v)-g(-u')}{g(-u)-g(-u')}\Big].\nn
\eea
Summing over all the terms one can find the relation Eq.(\ref{linearcombination}).
\end{widetext}


\begin{thebibliography}{999}

\bibitem{Amico:2007ag}
L.~Amico, R.~Fazio, A.~Osterloh and V.~Vedral, \textit{{Entanglement in
  many-body systems}}, \href{http://dx.doi.org/10.1103/RevModPhys.80.517}{Rev.
  Mod. Phys. {\bfseries 80}, 517 (2008)},
  [\href{https://arxiv.org/abs/quant-ph/0703044}{{\ttfamily
  arXiv:quant-ph/0703044}}].


\bibitem{Calabrese:2009qy}
P.~Calabrese and J.~Cardy, \textit{{Entanglement entropy and conformal field
  theory}}, \href{http://dx.doi.org/10.1088/1751-8113/42/50/504005}{J. Phys. A:
  Math. Gen. {\bfseries 42}, 504005 (2009)},
  [\href{https://arxiv.org/abs/0905.4013}{{\ttfamily arXiv:0905.4013}}].

\bibitem{Rangamani:2016dms}
M.~Rangamani and T.~Takayanagi, \textit{{Holographic Entanglement Entropy}},
  \href{http://dx.doi.org/10.1007/978-3-319-52573-0}{Lect. Notes Phys.
  {\bfseries 931}, 1--246 (2017)},
  [\href{https://arxiv.org/abs/1609.01287}{{\ttfamily arXiv:1609.01287}}].





\bibitem{Ryu:2006bv}
S.~Ryu and T.~Takayanagi, \textit{{Holographic Derivation of Entanglement
  Entropy from the anti-de Sitter Space/Conformal Field Theory
  Correspondence}},
  \href{http://dx.doi.org/10.1103/PhysRevLett.96.181602}{Phys. Rev. Lett.
  {\bfseries 96}, 181602 (2006)},
  [\href{https://arxiv.org/abs/hep-th/0603001}{{\ttfamily
  arXiv:hep-th/0603001}}].

\bibitem{Hubeny:2007xt}
V.~E. Hubeny, M.~Rangamani and T.~Takayanagi, \textit{{A Covariant holographic
  entanglement entropy proposal}},
  \href{http://dx.doi.org/10.1088/1126-6708/2007/07/062}{JHEP {\bfseries 07}
  (2007) 062}, [\href{https://arxiv.org/abs/0705.0016}{{\ttfamily
  arXiv:0705.0016}}].

\bibitem{VanRaamsdonk:2010pw}
M.~Van~Raamsdonk, \textit{{Building up spacetime with quantum entanglement}},
  \href{http://dx.doi.org/10.1007/s10714-010-1034-0}{Gen. Rel. Grav. {\bfseries
  42}, 2323 (2010)}, [\href{https://arxiv.org/abs/1005.3035}{{\ttfamily
  arXiv:1005.3035}}]. [\href{http://dx.doi.org/10.1142/S0218271810018529}
  {\textit{Int. J. Mod. Phys.} {\bfseries D19} (2010) 2429}].

\bibitem{Maldacena:2013xja}
J.~Maldacena and L.~Susskind,
\textit{Cool horizons for entangled black holes},
  \href{http://dx.doi.org/10.1002/prop.201300020}{Fortsch. Phys. \textbf{61} (2013), 781-811}, [\href{https://arxiv.org/abs/1306.0533}{{\ttfamily arXiv:1306.0533}}].


\bibitem{Almheiri:2014lwa}
A.~Almheiri, X.~Dong and D.~Harlow, \textit{{Bulk Locality and Quantum Error
  Correction in AdS/CFT}},
  \href{http://dx.doi.org/10.1007/JHEP04(2015)163}{JHEP {\bfseries 04} (2015)
  163}, [\href{https://arxiv.org/abs/1411.7041}{{\ttfamily arXiv:1411.7041}}].

\bibitem{Penington:2019npb}
G.~Penington, \textit{{Entanglement Wedge Reconstruction and the Information
  Paradox}}, \href{http://dx.doi.org/10.1007/JHEP09(2020)002}{JHEP {\bfseries
  09} (2020) 002}, [\href{https://arxiv.org/abs/1905.08255}{{\ttfamily
  arXiv:1905.08255}}].

\bibitem{Almheiri:2019psf}
A.~Almheiri, N.~Engelhardt, D.~Marolf and H.~Maxfield, \textit{{The entropy of
  bulk quantum fields and the entanglement wedge of an evaporating black
  hole}}, \href{http://dx.doi.org/10.1007/JHEP12(2019)063}{JHEP {\bfseries 12}
  (2019) 063}, [\href{https://arxiv.org/abs/1905.08762}{{\ttfamily
  arXiv:1905.08762}}].




\bibitem{Doi:2022iyj}
K.~Doi, J.~Harper, A.~Mollabashi, T.~Takayanagi and Y.~Taki,
  \textit{{Pseudoentropy in dS/CFT and Timelike Entanglement Entropy}},
  \href{http://dx.doi.org/10.1103/PhysRevLett.130.031601}{Phys. Rev. Lett.
  {\bfseries 130}, 031601 (2023)},
  [\href{https://arxiv.org/abs/2210.09457}{{\ttfamily arXiv:2210.09457}}].

\bibitem{Leggett:1985zz}
A.~J.~Leggett and A.~Garg,
\textit{{Quantum mechanics versus macroscopic realism: Is the flux there when nobody looks?,}},
  \href{http://dx.doi.org/10.1103/PhysRevLett.54.857}{Phys. Rev. Lett. \textbf{54} (1985), 857-860}.

\bibitem{Fitzsimons:2013gga}
J.~Fitzsimons, J.~Jones and V.~Vedral,
\textit{Quantum correlations which imply causation}, \href{https://doi.org/10.1038/srep18281}{Sci Rep 5, 18281 (2016). },
  [\href{https://arxiv.org/abs/1302.2731}{{\ttfamily arXiv:1302.2731 [quant-ph]}}].

\bibitem{Olson:2010jy}
S.~J.~Olson and T.~C.~Ralph,
\textit{Entanglement between the future and past in the quantum vacuum,}
\href{http://dx.doi:10.1103/PhysRevLett.106.110404}{Phys. Rev. Lett. \textbf{106} (2011), 110404},
  [\href{https://arxiv.org/abs/1003.0720}{{\ttfamily arXiv:1003.0720 [quant-ph]}}].

\bibitem{Giudice:2021smd}
G.~Giudice, G.~Giudici, M.~Sonner, J.~Thoenniss, A.~Lerose, D.~A.~Abanin and L.~Piroli,
\textit{Temporal Entanglement, Quasiparticles, and the Role of Interactions},
 \href{http://dx.doi.org/10.1103/PhysRevLett.128.220401}{Phys. Rev. Lett. \textbf{128} (2022) no.22, 220401},
  [\href{https://arxiv.org/abs/2112.14264}{{\ttfamily arXiv:2112.14264 [cond-mat.stat-mech]}}].


\bibitem{Liu:2022ugc}
B.~Liu, H.~Chen and B.~Lian,
\textit{Entanglement Entropy of Free Fermions in Timelike Slices,}
\href{https://arxiv.org/abs/2210.03134}{{\ttfamily arXiv:2210.03134 [cond-mat.stat-mech]}}.




\bibitem{Nakata:2020luh}
Y.~Nakata, T.~Takayanagi, Y.~Taki, K.~Tamaoka and Z.~Wei, \textit{{New
  holographic generalization of entanglement entropy}},
  \href{http://dx.doi.org/10.1103/PhysRevD.103.026005}{Phys. Rev. D {\bfseries
  103}, 026005 (2021)}, [\href{https://arxiv.org/abs/2005.13801}{{\ttfamily
  arXiv:2005.13801}}].

\bibitem{Murciano:2021dga}
S.~Murciano, P.~Calabrese and R.~M. Konik, \textit{{Generalized entanglement
  entropies in two-dimensional conformal field theory}},
  \href{http://dx.doi.org/10.1007/JHEP05(2022)152}{JHEP {\bfseries 05} (2022)
  152}, [\href{https://arxiv.org/abs/2112.09000}{{\ttfamily
  arXiv:2112.09000}}].

\bibitem{Mollabashi:2020yie}
A.~Mollabashi, N.~Shiba, T.~Takayanagi, K.~Tamaoka and Z.~Wei, \textit{{Pseudo
  Entropy in Free Quantum Field Theories}},
  \href{http://dx.doi.org/10.1103/PhysRevLett.126.081601}{Phys. Rev. Lett.
  {\bfseries 126}, 081601 (2021)},
  [\href{https://arxiv.org/abs/2011.09648}{{\ttfamily arXiv:2011.09648}}].


\bibitem{Guo:2022jzs}
W.-z. Guo, S.~He and Y.-X. Zhang, \textit{{Constructible reality condition of
  pseudoentropy via pseudo-Hermiticity}},
  \href{http://dx.doi.org/10.1007/JHEP05(2023)021}{JHEP {\bfseries 05} (2023)
  021}, [\href{https://arxiv.org/abs/2209.07308}{{\ttfamily
  arXiv:2209.07308}}].





\bibitem{Mollabashi:2021xsd}
A.~Mollabashi, N.~Shiba, T.~Takayanagi, K.~Tamaoka and Z.~Wei, \textit{{Aspects
  of pseudoentropy in field theories}},
  \href{http://dx.doi.org/10.1103/PhysRevResearch.3.033254}{Phys. Rev. Res.
  {\bfseries 3}, 033254 (2021)},
  [\href{https://arxiv.org/abs/2106.03118}{{\ttfamily arXiv:2106.03118}}].

\bibitem{Nishioka:2021cxe}
T.~Nishioka, T.~Takayanagi and Y.~Taki, \textit{{Topological pseudoentropy}},
  \href{http://dx.doi.org/10.1007/JHEP09(2021)015}{JHEP {\bfseries 09} (2021)
  015}, [\href{https://arxiv.org/abs/2107.01797}{{\ttfamily
  arXiv:2107.01797}}].

\bibitem{Goto:2021kln}
K.~Goto, M.~Nozaki and K.~Tamaoka, \textit{{Subregion spectrum form factor via
  pseudoentropy}}, \href{http://dx.doi.org/10.1103/PhysRevD.104.L121902}{Phys.
  Rev. D {\bfseries 104}, L121902 (2021)},
  [\href{https://arxiv.org/abs/2109.00372}{{\ttfamily arXiv:2109.00372}}].

\bibitem{Akal:2021dqt}
I.~Akal, T.~Kawamoto, S.-M. Ruan, T.~Takayanagi and Z.~Wei, \textit{{Page curve
  under final state projection}},
  \href{http://dx.doi.org/10.1103/PhysRevD.105.126026}{Phys. Rev. D {\bfseries
  105}, 126026 (2022)}, [\href{https://arxiv.org/abs/2112.08433}{{\ttfamily
  arXiv:2112.08433}}].

\bibitem{Miyaji:2021lcq}
M.~Miyaji,
\textit{Island for gravitationally prepared state and pseudo entanglement wedge},
\href{http://dx.doi.org/10.1007/JHEP12(2021)013}{
JHEP \textbf{12}, 013 (2021)},
[\href{https://arxiv.org/abs/2109.03830}{{\ttfamily arXiv:2109.03830}}].



\bibitem{Guo:2022sfl}
W.-z. Guo, S.~He and Y.-X. Zhang, \textit{{On the real-time evolution of
  pseudo-entropy in 2d CFTs}},
  \href{http://dx.doi.org/10.1007/JHEP09(2022)094}{JHEP {\bfseries 09} (2022)
  094}, [\href{https://arxiv.org/abs/2206.11818}{{\ttfamily arXiv:2206.11818}}].

\bibitem{Ishiyama:2022odv}
Y.~Ishiyama, R.~Kojima, S.~Matsui and K.~Tamaoka,
\textit{Notes on pseudoentropy amplification},
\href{http://dx.doi.org/10.1093/ptep/ptac112}{PTEP \textbf{2022}, no.9, 093B10 (2022)},
[\href{https://arxiv.org/abs/2206.14551}{{\ttfamily arXiv:2206.14551}}]
\bibitem{Mukherjee:2022jac}
J.~Mukherjee,
\textit{Pseudo Entropy in U(1) gauge theory},
\href{https://link.springer.com/article/10.1007/JHEP10(2022)016}{JHEP \textbf{10}(2022)016},
[\href{https://arxiv.org/abs/2205.08179}{\ttfamily{arXiv:2205.08179}}].

\bibitem{Li:2022tsv}
Z.~Li, Z.-Q.~Xiao and R.-Q.~Yang, \textit{{On holographic time-like entanglement entropy}}, \href{http://dx.doi.org/10.1007/JHEP04(2023)004}{JHEP {\bfseries 04} (2023) 004}, [\href{https://arxiv.org/abs/2211.14883}{{\ttfamily arXiv:2211.14883}}].

\bibitem{He:2023eap}
S.~He, J.~Yang, Y.-X.~Zhang and Z.-X.~Zhao, \textit{{Pseudo-entropy for descendant operators in two-dimensional conformal field theories}}, [\href{https://arxiv.org/abs/2301.04891}{{\ttfamily arXiv:2301.04891}}].

\bibitem{Narayan:2022afv}
K.~Narayan, \textit{{de Sitter space, extremal surfaces, and time entanglement}}, \href{http://dx.doi.org/10.1103/PhysRevD.107.126004}{Phys. Rev. D {\bfseries 107} (2023) 126004}, [\href{https://arxiv.org/abs/2210.12963}{{\ttfamily arXiv:2210.12963}}].

\bibitem{Doi:2023zaf}
K.~Doi, J.~Harper, A.~Mollabashi, T.~Takayanagi and Y.~Taki, \textit{{Timelike entanglement entropy}}, \href{http://dx.doi.org/10.1007/JHEP05(2023)052}{JHEP {\bfseries 05} (2023) 052}, [\href{https://arxiv.org/abs/2302.11695}{{\ttfamily arXiv:2302.11695}}].

\bibitem{Kawamoto:2023nki}
T.~Kawamoto, S.~M.~Ruan, Y.-k.~Suzuki and T.~Takayanagi, \textit{{A half de Sitter holography}}, \href{https://link.springer.com/article/10.1007/JHEP10(2023)137}{JHEP {\bfseries 10} (2023) 137}, [\href{https://arxiv.org/abs/2306.07575}{{\ttfamily arXiv:2306.07575}}].

\bibitem{Narayan:2023ebn}
K.~Narayan and H.~K.~Saini, \textit{{Notes on time entanglement and pseudo-entropy}}, [\href{https://arxiv.org/abs/2303.01307}{{\ttfamily arXiv:2303.01307}}].

\bibitem{He:2023wko}
S.~He, J.~Yang, Y.~X.~Zhang and Z.~X.~Zhao, \textit{{Pseudo entropy of primary operators in $T\bar{T}$/$J\bar{T}$-deformed CFTs}}, [\href{https://arxiv.org/abs/2305.10984}{{\ttfamily arXiv:2305.10984}}].

\bibitem{Jiang:2023ffu}
X.~Jiang, P.~Wang, H.~Wu and H.~Yang, \textit{{Timelike entanglement entropy and $T\bar{T}$ deformation}}, [\href{https://arxiv.org/abs/2302.13872}{{\ttfamily arXiv:2302.13872}}].

\bibitem{Parzygnat:2023avh}
A.~J.~Parzygnat, T.~Takayanagi, Y.~Taki and Z.~Wei, \textit{{SVD Entanglement Entropy}}, [\href{https://arxiv.org/abs/2307.06531}{{\ttfamily arXiv:2307.06531}}].

\bibitem{Guo:2023aio}
W.-z.~Guo and J.~Zhang, \textit{{Sum rule for pseudo-R\'enyi entropy}}, [\href{https://arxiv.org/abs/2308.05261}{{\ttfamily arXiv:2308.05261}}].

\bibitem{He:2023syy}
S.~He, Y.~X.~Zhang, L.~Zhao and Z.~X.~Zhao, \textit{{Entanglement and Pseudo Entanglement Dynamics versus Fusion in CFT}}, [\href{https://arxiv.org/abs/2312.02679}{{\ttfamily arXiv:2312.02679}}].

\bibitem{Das:2023yyl}
A.~Das, S.~Sachdeva and D.~Sarkar, \textit{{Bulk reconstruction using timelike entanglement in (A)dS}}, [\href{https://arxiv.org/abs/2312.16056}{{\ttfamily arXiv:2312.16056}}].

\bibitem{Chu:2023zah}
C.~S.~Chu and H.~Parihar, \textit{{Time-like entanglement entropy in AdS/BCFT}}, \href{https://link.springer.com/article/10.1007/JHEP06(2023)173}{JHEP {\bfseries 06} (2023) 173}, [\href{https://arxiv.org/abs/2304.10907}{{\ttfamily arXiv:2304.10907}}].

\bibitem{Jiang:2023loq}
X.~Jiang, P.~Wang, H.~Wu and H.~Yang, \textit{{Timelike entanglement entropy in dS$_{3}$/CFT$_{2}$}}, \href{https://link.springer.com/article/10.1007/JHEP08(2023)216}{JHEP {\bfseries 08} (2023) 216}, [\href{https://arxiv.org/abs/2304.10376}{{\ttfamily arXiv:2304.10376}}].

\bibitem{Chen:2023gnh}
Z.~Chen, \textit{{Complex-valued Holographic Pseudo Entropy via Real-time AdS/CFT Correspondence}}, [\href{https://arxiv.org/abs/2302.14303}{{\ttfamily arXiv:2302.14303}}].

\bibitem{He:2023ubi}
P.~Z.~He and H.~Q.~Zhang, \textit{{Timelike Entanglement Entropy from Rindler Method}}, [\href{https://arxiv.org/abs/2307.09803}{{\ttfamily arXiv:2307.09803}}].

\bibitem{Omidi:2023env}
F.~Omidi, \textit{{Pseudo R\'enyi Entanglement Entropies For an Excited State and Its Time Evolution in a 2D CFT}}, [\href{https://arxiv.org/abs/2309.04112}{{\ttfamily arXiv:2309.04112}}].

\bibitem{Narayan:2023zen}
K.~Narayan, \textit{{Further remarks on de Sitter space, extremal surfaces and time entanglement}}, [\href{https://arxiv.org/abs/2310.00320}{{\ttfamily arXiv:2310.00320}}].

\bibitem{Guo:2023tjv}
W.-z.~Guo and Y.~Jiang, \textit{{Pseudo entropy and pseudo-Hermiticity in quantum field theories}}, [\href{https://arxiv.org/abs/2311.01045}{{\ttfamily arXiv:2311.01045}}].

\bibitem{Kanda:2023jyi}
H.~Kanda, T.~Kawamoto, Y.-k.~Suzuki, T.~Takayanagi, K.~Tasuki and Z.~Wei, \textit{{Entanglement Phase Transition in Holographic Pseudo Entropy}}, [\href{https://arxiv.org/abs/2311.13201}{{\ttfamily arXiv:2311.13201}}].

\bibitem{Shinmyo:2023eci}
K.~Shinmyo, T.~Takayanagi and K.~Tasuki, \textit{{Pseudo entropy under joining local quenches}}, [\href{https://arxiv.org/abs/2310.12542}{{\ttfamily arXiv:2310.12542}}].

\bibitem{Heller:2024whi}
M.~P.~Heller, F.~Ori and A.~Serantes, \textit{{Geometric Interpretation of Timelike Entanglement Entropy}}, \href{https://doi.org/10.1103/PhysRevLett.134.131601}{Phys. Rev. Lett. {\bfseries 134} (2025) 131601}, [\href{https://arxiv.org/abs/2408.15752}{{\ttfamily arXiv:2408.15752}}].

\bibitem{Guo:2024ITE}
W.-z.~Guo and J.~Xu, \textit{{Imaginary part of timelike entanglement entropy}}, \href{https://doi.org/10.1007/JHEP02(2025)094}{JHEP {\bfseries 02} (2025) 094}, [\href{https://arxiv.org/abs/2410.22684}{{\ttfamily arXiv:2410.22684}}].

\bibitem{Guo:2025pru}
W.-z.~Guo and J.~Xu, \textit{{A duality of Ryu-Takayanagi surfaces inside and outside the horizon}}, [\href{https://arxiv.org/abs/2502.16774}{{\ttfamily arXiv:2502.16774}}].

\bibitem{Milekhin:2025ycm}
A.~Milekhin, Z.~Adamska and J.~Preskill, \textit{{Observable and computable entanglement in time}}, [\href{https://arxiv.org/abs/2502.12240}{{\ttfamily arXiv:2502.12240}}].

\bibitem{Nunez:2025ppd}
C.~Nunez and D.~Roychowdhury, \textit{{Interpolating between Space-like and Time-like Entanglement via Holography}}, [\href{https://arxiv.org/abs/2507.17805}{{\ttfamily arXiv:2507.17805}}].

\bibitem{Heller:2025kvp}
M.~P.~Heller, F.~Ori and A.~Serantes, \textit{{Temporal Entanglement from Holographic Entanglement Entropy}}, [\href{https://arxiv.org/abs/2507.17847}{{\ttfamily arXiv:2507.17847}}].

\bibitem{Gong:2025pnu}
X.~Gong, W.-z.~Guo and J.~Xu, \textit{{Entanglement measures for causally connected subregions and holography}}, [\href{https://arxiv.org/abs/2508.05158}{{\ttfamily arXiv:2508.05158}}].

\bibitem{Calabrese:2004eu}
P.~Calabrese and J.~L.~Cardy, \textit{{Entanglement entropy and quantum field theory}}, \href{http://dx.doi.org/10.1088/1742-5468/2004/06/P06002}{J. Stat. Mech. (2004) P06002}, [\href{https://arxiv.org/abs/hep-th/0405152}{{\ttfamily arXiv:hep-th/0405152}}].

\bibitem{Guo:2023tys}
W.-z.~Guo and J.~Xu, \textit{{Parameter dependence of entanglement spectra in quantum field theories}}, [\href{https://arxiv.org/abs/2312.13688}{{\ttfamily arXiv:2312.13688}}].

\bibitem{Borchers}
H.~J.~Borchers, \textit{{Über die Vollständigkeit Lorentzinvarianter Felder in einer zeitartigen Röhre}}, Nuovo Cimento {\bfseries 19} (1961) 787.

\bibitem{Araki}
H.~Araki, \textit{{A Generalization Of Borchers' Theorem}}, Helv. Phys. Acta {\bfseries 36} (1963) 132.

\bibitem{Strohmaier:2023hhy}
A.~Strohmaier and E.~Witten, \textit{{Analytic states in quantum field theory on curved spacetimes}}, [\href{https://arxiv.org/abs/2302.02709}{{\ttfamily arXiv:2302.02709}}].

\bibitem{Strohmaier:2023opz}
A.~Strohmaier and E.~Witten, \textit{{The Timelike Tube Theorem in Curved Spacetime}}, [\href{https://arxiv.org/abs/2303.16380}{{\ttfamily arXiv:2303.16380}}].

\bibitem{Chandrasekaran:2022cip}
V.~Chandrasekaran, R.~Longo, G.~Penington and E.~Witten, \textit{{An algebra of observables for de Sitter space}}, \href{https://link.springer.com/article/10.1007/JHEP02(2023)082}{JHEP {\bfseries 02} (2023) 082}, [\href{https://arxiv.org/abs/2206.10780}{{\ttfamily arXiv:2206.10780}}].

\bibitem{Chandrasekaran:2022eqq}
V.~Chandrasekaran, G.~Penington and E.~Witten, \textit{{Large N algebras and generalized entropy}}, \href{https://link.springer.com/article/10.1007/JHEP04(2023)009}{JHEP {\bfseries 04} (2023) 009}, [\href{https://arxiv.org/abs/2209.10454}{{\ttfamily arXiv:2209.10454}}].

\bibitem{Witten:2023qsv}
E.~Witten, \textit{{Algebras, regions, and observers}}, Proc. Symp. Pure Math. {\bfseries 107} (2024) 247, [\href{https://arxiv.org/abs/2303.02837}{{\ttfamily arXiv:2303.02837}}].

\bibitem{Witten:2023xze}
E.~Witten, \textit{{A background-independent algebra in quantum gravity}}, \href{https://link.springer.com/article/10.1007/JHEP03(2024)077}{JHEP {\bfseries 03} (2024) 077}, [\href{https://arxiv.org/abs/2308.03663}{{\ttfamily arXiv:2308.03663}}].

\bibitem{Cardy:2007mb}
J.~L.~Cardy, O.~A.~Castro-Alvaredo and B.~Doyon, \textit{{Form factors of branch-point twist fields in quantum integrable models and entanglement entropy}}, \href{http://dx.doi.org/10.1007/s10955-007-9422-x}{J. Stat. Phys. {\bfseries 130} (2008) 129}, [\href{https://arxiv.org/abs/0706.3384}{{\ttfamily arXiv:0706.3384}}].

\bibitem{Headrick:2010zt}
M.~Headrick, \textit{{Entanglement R\'enyi entropies in holographic theories}}, \href{http://dx.doi.org/10.1103/PhysRevD.82.126010}{Phys. Rev. D {\bfseries 82} (2010) 126010}, [\href{https://arxiv.org/abs/1006.0047}{{\ttfamily arXiv:1006.0047}}].


\bibitem{Calabrese:2010he}
P.~Calabrese, J.~Cardy and E.~Tonni, \textit{{Entanglement entropy of two disjoint intervals in conformal field theory II}},
\href{http://dx.doi.org/10.1088/1742-5468/2011/01/P01021}{J. Stat. Mech. (2011) P01021},
[\href{https://arxiv.org/abs/1011.5482}{{\ttfamily arXiv:1011.5482}}].

\bibitem{Rajabpour:2011pt}
M.~A.~Rajabpour and F.~Gliozzi, \textit{{Entanglement Entropy of Two Disjoint Intervals from Fusion Algebra of Twist Fields}},
\href{http://dx.doi.org/10.1088/1742-5468/2012/02/P02016}{J. Stat. Mech. (2012) P02016},
[\href{https://arxiv.org/abs/1112.1225}{{\ttfamily arXiv:1112.1225}}].

\bibitem{Chen:2013kpa}
B.~Chen and J.-j.~Zhang, \textit{{On short interval expansion of R\'enyi entropy}},
\href{http://dx.doi.org/10.1007/JHEP11(2013)164}{JHEP {\bfseries 11} (2013) 164},
[\href{https://arxiv.org/abs/1309.5453}{{\ttfamily arXiv:1309.5453}}].

\bibitem{Hartman:2014oaa}
T.~Hartman, C.~A.~Keller and B.~Stoica, \textit{{Universal Spectrum of 2d Conformal Field Theory in the Large c Limit}},
\href{http://dx.doi.org/10.1007/JHEP09(2014)118}{JHEP {\bfseries 09} (2014) 118},
[\href{https://arxiv.org/abs/1405.5137}{{\ttfamily arXiv:1405.5137}}].

\bibitem{Asplund:2014coa}
C.~T.~Asplund, A.~Bernamonti, F.~Galli and T.~Hartman, \textit{{Holographic Entanglement Entropy from 2d CFT: Heavy States and Local Quenches}},
\href{http://dx.doi.org/10.1007/JHEP02(2015)171}{JHEP {\bfseries 02} (2015) 171},
[\href{https://arxiv.org/abs/1410.1392}{{\ttfamily arXiv:1410.1392}}].

\bibitem{Casini:2011kv}
H.~Casini, M.~Huerta and R.~C.~Myers, \textit{{Towards a derivation of holographic entanglement entropy}},
\href{http://dx.doi.org/10.1007/JHEP05(2011)036}{JHEP {\bfseries 05} (2011) 036},
[\href{https://arxiv.org/abs/1102.0440}{{\ttfamily arXiv:1102.0440}}].

\bibitem{Strominger:2001pn}
A.~Strominger, \textit{{The dS / CFT correspondence}},
\href{http://dx.doi.org/10.1088/1126-6708/2001/10/034}{JHEP {\bfseries 10} (2001) 034},
[\href{https://arxiv.org/abs/hep-th/0106113}{{\ttfamily arXiv:hep-th/0106113}}].






\end{thebibliography}


\end{document}